\documentclass[reprint,superscriptaddress,
 amsmath,amssymb,
 aps,prb
floatfix,
]{revtex4-2}

\RequirePackage{hyperref}

\usepackage{dcolumn}
\usepackage{bm}
\usepackage{gensymb}
\usepackage[dvipsnames]{xcolor}

\usepackage[final]{spin-textures}
\RequirePackage{mathtools}

\graphicspath{{fig}}

\begin{document}
\doclicenseThis 


\title{Non-reciprocity and exchange-spring delay of domain-wall Walker breakdown in magnetic nanowires with azimuthal magnetization}%

\author{Luc\'ia G\'omez-Cruz\,\orcidlink{0009-0006-9777-429X}}
\affiliation{Dpto. F\'isica de Materiales, Universidad Complutense de Madrid, Plaza de las Ciencias 1, Madrid, 28040, Spain}
\affiliation{Univ. Grenoble Alpes, CNRS, CEA, SPINTEC, Grenoble INP, 38000 Grenoble, France.}

\author{Laura \'Alvaro-G\'omez\,\orcidlink{0000-0001-8899-3518}}
 \affiliation{Dpto. F\'isica de Materiales, Universidad Complutense de Madrid, Plaza de las Ciencias 1, Madrid, 28040, Spain}

\author{Claudia Fern\'andez-Gonz\'alez\,\orcidlink{0000-0002-6299-5803}}
\author{Sandra Ruiz-G\'omez\,\orcidlink{0000-0002-9665-2059}}
\affiliation{Alba Synchrotron Light Facility, CELLS, Carrer de la Llum 2-26, Cerdanyola del Vall\`es, 08290, Barcelona, Spain}

\author{Christophe Thirion\,\orcidlink{0000-0001-6400-7971}}
\affiliation{Univ. Grenoble Alpes, CNRS, Institut Néel, 38000 Grenoble, France}

\author{Giuseppe Curci\,\orcidlink{0009-0004-1764-8134}}
\affiliation{Univ. Grenoble Alpes, CNRS, CEA, SPINTEC, 38000 Grenoble, France.}

\author{Lucia Aballe\,\orcidlink{0000-0003-1810-8768}}
\author{Eva Pereiro\,\orcidlink{0000-0001-7626-5935}}
\affiliation{Alba Synchrotron Light Facility, CELLS, Carrer de la Llum 2-26, Cerdanyola del Vall\`es, 08290, Barcelona, Spain}

\author{Rachid Belkhou\,\orcidlink{0000-0002-2218-7481}}
\affiliation{Synchrotron SOLEIL, Saint-Aubin, F-91192 Gif-sur-Yvette, France}

\author{Eduardo Mart\'\i nez\,\orcidlink{0000-0003-2960-5508}}
\author{Victor Raposo\,\orcidlink{/0000-0002-6892-036X}}
\affiliation{Department of Applied Physics, Universidad de Salamanca, E-37008 Salamanca, Spain}

\author{Jean-Christophe Toussaint\,\orcidlink{0000-0001-5382-3594}}
\affiliation{Univ. Grenoble Alpes, CNRS, Institut Néel, 38000 Grenoble, France}

\author{Daria Gusakova\,\orcidlink{https://orcid.org/0000-0002-5111-7079}}
\author{Aurélien~Masseboeuf\,\orcidlink{0000-0003-4239-1313}}
\author{Olivier~Fruchart\,\orcidlink{0000-0001-7717-5229}}
\affiliation{Univ. Grenoble Alpes, CNRS, CEA, SPINTEC, 38000 Grenoble, France.}

\author{Lucas~Pérez\,\orcidlink{0000-0001-9470-7987}}
 \affiliation{Dpto. F\'isica de Materiales, Universidad Complutense de Madrid, Plaza de las Ciencias 1, Madrid, 28040, Spain}
 \affiliation{Instituto Madrile\~no de Estudios Avanzados - IMDEA Nanociencia, C/ Faraday 9, Madrid, 28049, Spain}

\begin{abstract}
Domain-wall~(DW) motion is a crucial process involved in magnetization reversal, be it under magnetic field or spin-polarized current stimulus. In most cases \refereeAdd{the }DW speed does not exceed $\approx\qty{100}{\meter\per\second}$ and collapses above a given threshold of the stimulus, an effect known as Walker breakdown. A few specific material properties have been identified to delay the breakdown of speed by increasing the energy barrier preventing internal precession. We show that in a 3D nanomagnetic system, here with vortex-state domains, the topology of the magnetization distribution may intrinsically and robustly delay the Walker breakdown due to an exchange-spring effect. \refereeAdd{However, in contrast to situations in lower dimension, the motion remains steady-state}. In addition, curvature induces a major non-reciprocal effect, delaying or not the Walker breakdown depending on the \refereeAdd{sequence of }chiralit\refereeReplace{y}{ies} of the azimuthal domain\refereeAdd{s on both sides of the DW,} \refereeReplace{versus the}{independent of its} direction of motion\refereeRemove{ of the DW}.
\end{abstract}

\maketitle

\section{Introduction}

The response of a system to an external driving force may be linear at small drive amplitude and enter a turbulent regime above a critical threshold. A typical example is fluid dynamics, transiting from laminar to  turbulent flow for Reynolds numbers above a critical value, which depends on fluid inertia\refereeAdd{ and viscosity}, system dimension\refereeAdd{s}\refereeReplace{,}{ and} flow speed\refereeRemove{ and viscosity}. Domain wall~(DW) motion in a magnetic material shows an analoguous behavior, when set in motion by a magnetic field\cite{bib-THI2006} or a spin-polarized current\cite{bib-THI2008}: DWs usually display a constant \refereeAdd{and high }mobility at low drive, and enter a turbulent regime \refereeAdd{with low speed }above a so-called Walker breakdown field~(resp., current). The underlying physics is very rich, involving the continuous precession of magnetization inside the DW and possibly the nucleation of sub-textures in the magnetization of the DW such as vortices. Besides fundamental interest, the understanding and control of DW motion is crucial to develop reliable and efficient applications in data storage and logic\cite{bib-THO2007}.

Flat strips are a textbook situation to investigate DW dynamics\cite{bib-MOU2007}, owing to their nearly one-dimensional shape and thus limited number of degrees of freedom. The Walker breakdown generally limits speed to around \qty{100}{\meter\per\second} experimentally, such as in the prototypical soft magnetic material Permalloy, under either magnetic field\cite{bib-BEA2005} or electric current\cite{bib-HAY2007b,bib-HAY2007}. Some intrinsic material properties may delay the onset of DW precession, such as the Dzyaloshinskii-Moriya interaction~(DMI) promoting a given chirality of the DW\cite{bib-THI2012,bib-THI2012}, or the vanishing magnetization in compensated ferrimagnets\cite{bib-KIM2017c,bib-CAR2018}. This, however, remains material-specific.

Three-dimensional~(3D) magnetic systems have been receiving attention in the past decade\cite{bib-FRU2017b,bib-STR2021,bib-MAK2022}, as their shape can induce a number of effects, adding to the material intrinsic properties: curvature-induced magnetic anisotropy or DMI, chirality, and/or topology constraints, \eg\refereeAdd{,} the magnetic M\oe bius string \refereeAdd{with perpendicular magnetization,}preventing the existence of a uniform magnetic state\refereeAdd{ in the curvilinear coordinates}\cite{bib-PYL2015}. As regards DW motion, cylindrical structures are the textbook counterpart of flat strips for 3D systems. In some of these systems it has been predicted that the enhanced strength of magnetostatic energy in 3D delays the Walker breakdown by increasing the energy barrier associated with DW precession. This concerns head-to-head DWs in both cylindrical wires\cite{bib-THI2006,bib-HER2016}, the so-called Bloch DW, and in cylindrical nanotubes\cite{bib-YAN2011b}. As a consequence, speeds exceeding \qty{500}{\meter\per\second} have been predicted, with encouraging experimental reports\cite{bib-FRU2019b,bib-BRA2023}. \refereeAdd{Also, curvature has been shown to induce non-reciprocal effects in the propagation of spin waves for nanotubes with azimuthal magnetization\cite{bib-OTA2016,bib-KOE2022}. In these, spin waves have the Damon-Eshbach geometry\cite{bib-STA2009}, which is defined when the propagation direction $\vect k$ and magnetization $\vect M$ are perpendicular, in the present case along the axis and along the azimuthal direction, respectively. In contrast with thin films for which the top and bottom surface are equivalent, curvature lifts the degeneracy between the outer and inner surfaces in a tube.}

Here, we demonstrate a\refereeAdd{nother} material-independent effect delaying the Walker breakdown, \refereeAdd{the energy barrier being enhanced thanks to exchange instead of magnetostatics, and} stemming from the topology of a three-dimension\refereeAdd{al} magnetization texture in cylindrical nanowires (NWs). We consider DWs separating opposite\refereeReplace{-}{ }azimuthally-magnetized domains in soft-magnetic NWs with a uniformly-axially-magnetized core, also called a vortex state\cite{bib-RUI2018,bib-FER2022}\bracketsubfigref{fig-structure-statics}{a}. This original geometry allows us to set the DWs in motion by the \OErsted field arising from a charge current flowing through the wire. We evidenced speed exceeding \qty{500}{\meter\per\second}, which we explain by a magnetic exchange\refereeReplace{-}{ }bias between the \refereeAdd{almost uniformly-magnetized }core and the DW\refereeReplace{s}{ parts} at the wire periphery, a phenomenon analogous to the one proposed for exchange-spring magnets\cite{bib-COE1988,bib-KNE1991,bib-SKO1993}. Unlike DMI, whose energy barrier peaks for a $\pi$ rotation of the DW but the system comes back to its ground state for $2\pi$ rotation of magnetization, here the core hinders further rotation and therefore causes a substantial modification and  delay of the Walker phenomenon. Besides, due to the curvature of the wire surface this phenomenon \refereeReplace{is}{proves to be} chiral, depending on the relative \refereeAdd{sequence of the }orientation of peripheral magnetization \refereeAdd{in the two domains around the DW, \ie, left/right or right/left, but independent of} \refereeRemove{with respect to}the direction of DW motion.

The manuscript is organized as follows. We first describe the preparation and static properties of the samples considered. We then report DW motion under nanosecond pulses of charge current, imaged using Transmission X-ray Microscopy~(TXM) and ptychography combined with X-ray Magnetic Circular Dichroism~(XMCD). We finally link the high DW speed measured with the three-dimensional topology of magnetization, with the help of both analytical modeling and micromagnetic simulation\refereeAdd{, and proceed to discussion}.

\section{Experiments}

\begin{figure}[h]
    \includegraphics{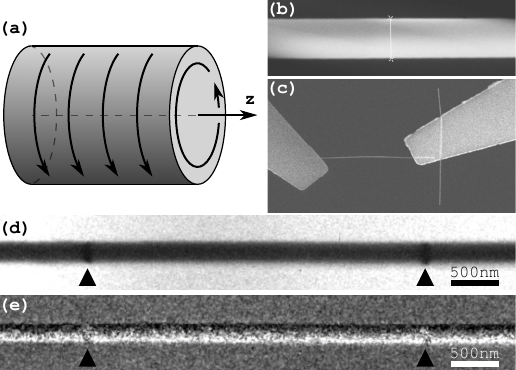}
    \caption{(a)~Sketch of a wire with azimuthal magnetization (b,c)~SEM images of a \qty{200}{\nano\meter}-diameter nanowire
    with Permalloy segments of average length \qty{3.6}{\micro\meter}, separated with \qty{40}{\nano\meter}-wide \ce{Fe_70Ni_30} chemical modulations: close-up and full view of an electrically-contacted nanowire. (d)~Transmission and XMCD-TXM views of such NWs, with two consecutive chemical modulations highlighted with black triangular marks.\dataref{ALBA Nov2023 R10 $144_\mathrm{pos}$ $-143_\mathrm{neg}$}}
    \label{fig-structure-statics}
\end{figure}

We fabricated \qty{200}{\nano\meter}-diameter cylindrical NWs by template-assisted electrodeposition\bracketfigref{fig-structure-statics}. The NWs are made of Permalloy segments, \ie, \ce{Fe_20Ni_80}, separated by \qty{40}{\nano\meter}-long chemical modulations of composition \refereeRemove{of }\ce{Fe_70Ni_30}. The template is dissolved and a diluted drop of wire-containing solution is let to dry on \qty{40}{\nano\meter}-thick \ce{Si_3N_4} windows held by a frame of Si wafer, to allow for transmission microscopy. Two electric pads for current pulses injection were subsequently patterned with laser lithography. We used TXM for magnetic imaging, with energy tuned to the Fe L$_3$ absorption edge\bracketsubfigref{fig-structure-statics}{d}. XMCD images were computed as the normalized difference of transmission images with left and right polarizations, which highlights the magnetization component collinear to the X-ray beam, either
parallel~(light) or antiparallel~(dark)\bracketsubfigref{fig-structure-statics}{e}\cite{bib-STO1998}. The dark/light contrast on opposite sides of the wire reveals an azimuthal component of magnetization at its periphery. Azimuthal magnetic domains curling around an axially-magnetized core had been reported already three decades ago in amorphous Co-rich magnetostrictive microwires\cite{bib-BUS1994}, with typical diameters of several tens of  \unit{\micro\meter}, and presenting the so-called Giant Magneto-Impedance effect\cite{bib-PAN1994}. Recently, azimuthal peripheral magnetic domains have also been reported in Permalloy nanowires with diameter of a few hundreds of nanometers\cite{bib-RUI2018,bib-FER2022}. While it was initially thought that azimuthal domains arise \refereeAdd{due to and} in the vicinity of \refereeAdd{larger-magnetization }chemical modulations\refereeAdd{ owing to their intrinsic tendancy for curling\cite{bib-FRU2022}}, it is now recognized that homogeneous wires \refereeAdd{even without chemical modulations but} of sufficient diameter may host them, typically above \qty{150}{\nano\meter}.

Hereafter, domains characterized by a light-top and dark-bottom contrast will be referred to as ($C+$), whose axial vector is parallel to~$+\vect z$, while the reverse will be labeled \refereeRemove{as }($C-$). The sign of the \OErsted field resulting from the pulses of current will be labeled the same, so that positive~(resp. negative) current pulses are expected to favor domains with right~(resp. left) circulation.

\begin{figure}[t]
    \includegraphics{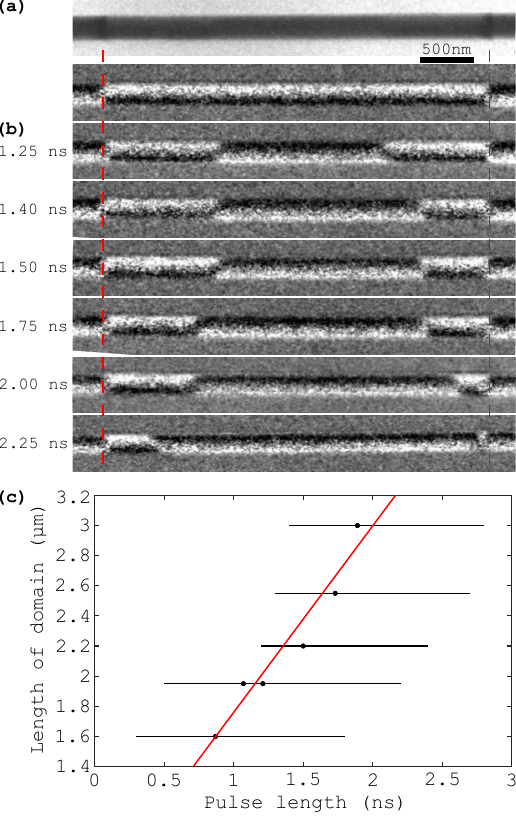}
    \caption{(a)~Transmission and XMCD image of the initial state. (b)~Starting from the same initial state, each row is a static XMCD image taken after the application of a current pulse of $J = \qty{1.1E11}{\ampere\per\meter\squared}$, with duration indicated on the left-hand side. After each step the initial state is recovered by applying a pulse with opposite polarity. The position of the chemical modulations is highlighted with vertical dashed red lines (c)~Length of the nucleated domain versus the length of the current pulse. The horizontal lines are error bars~(see End matter).}
    \label{fig-current-pulses}
\end{figure}

The bandwidth of the setup is a few hundreds of \unit{\mega\hertz}. It is limited by the pulse generator, not by the sample \refereeReplace{mounting}{leads and connectors}\refereeRemove{and oscilloscope}. Consequently, we cannot apply pulses of current shorter than \qty{500}{\pico\second}, and the varying current during the rise and fall times introduce a bias in the analysis\refereeAdd{ of the DW speed}~(see End Matter). Therefore, in order to precisely infer the DW speed, we did not image a given DW after successive pulses, but instead we reset the sample circulation with a first current pulse\refereeAdd{\bracketsubfigref{fig-current-pulses}a} and nucleate a fresh \refereeReplace{DW}{domain }with an opposite pulse in a reproducible fashion between every two static images. The length of the second pulse can be adjusted finely\refereeAdd{\bracketsubfigref{fig-current-pulses}b}, allowing us to monitor the DW displacement in intervals down to \qty{100}{\pico\second}, which is the nominal time resolution of our pulse generator\refereeAdd{, \ie, below the bandwidth of the pulse generator}. This approach also allows to disregard non-linear DW motion at the beginning and end of the pulses, such as inertia and threshold of current for motion. The way to define the pulse length is detailed in the End Matter section. \refereeRemove{The transmission image and the initial (reset) magnetic state are shown in \subfigref{fig-current-pulses}(a). }\refereeReplace{A}{For the moderate value of current used for the reset pulse, only the central segment is reset, so that the initial state consists of a} ($C+$) domain \refereeRemove{is} bonded by two domain walls pinned at the modulations. \subfigref{fig-current-pulses}(b)  shows the magnetic states resulting from the application of a pulse with amplitude $\qty{1.1E11}{\ampere\per\square\meter}$ and increasing duration. The longer the pulse is, the further the DWs have moved. This monotonous variation of DW position suggests that the nucleation site and process are reproducible, probably matching a specific chemical or structural defect in the nanowire. \refereeReplace{W}{Besides, w}e estimate\refereeAdd{d} a temperature rise, due to Joule heating during a pulse, to be one degree or less~(see End Matter). Thermal effects can thus be safely neglected.

\figref{fig-current-pulses}(c) displays the length of the nucleated domain versus pulse length. The \refereeRemove{nucleation} length \refereeRemove{is $\approx\qty{500}{\nano\meter}$, \ie,} extrapolated to null duration \refereeAdd{is $\approx\qty{500}{\nano\meter}$, \ie}, with some uncertainty due to the definition of pulse width and influence of DW inertia\refereeAdd{, and is related to the length of the nucleation domain}. The DW speed can be inferred to $\vDW \approx \qty{600}{\meter\per\second}$, averaging the speed of both DWs. This value comes with an uncertainty of the order of $\qty{100}{\meter\per\second}$, owing to the error bars related to the control of pulse width, and possible stochastic DW pinning along the wire. Note that we could not investigate speed over an extensive range of applied current in the present wires. Indeed a lower bound of usable current is set by the critical value for nucleation. Conversely, for large current we have measured DW speeds up to \qty{2}{\kilo\meter\per\second}. The latter values, however, had a large spread and depended on the pulse duration. Time-resolved imaging revealed that the mechanism is not DW motion but coherent switching of the circulation, occurring when the \OErsted field exceed the strength of azimuthal anisotropy~(See End Matter). Consequently, we restrict our analysis to the current density $\qty{1.1E11}{\ampere\per\meter\squared}$, which corresponds to an \OErsted field with magnitude \qty{7}{\milli\tesla} at the periphery of the wire. This is of the order of the Walker field in thin films made of a soft-magnetic material, typically \qtyrange{2}{3}{\milli\tesla}, which however are limited to speed around $\qty{100}{\meter\per\second}$. This suggests that a specific phenomenon delays the Walker breakdown in the present situation.

DW motion in microwires with azimuthal magnetization \refereeAdd{has been identified to }give\refereeRemove{s} rise to Giant Magneto-Impedance in amorphous microwires at frequencies below those of ferromagnetic resonance, induced by \refereeAdd{DW motion driven by }the \OErsted field associated to the charge current flowing in the wire\cite{bib-BUS1994}. It can also induce the full switching of the circulation\cite{bib-CHI2003}. The switching of peripheral circulation was imaged directly \cite{bib-CHI2009}, however only the speed of head-to-head domain walls arising on the wire axis was quantified\cite{bib-ZHU2018}. Nanowires, with far fewer degrees of freedom, offer the opportunity to investigate physics down to its basic ingredients. \refereeReplace{It will also be shown}{We will also show} that the much larger role of exchange in NWs \refereeAdd{related to their smaller size }plays a crucial role in DW motion.

\section{Analysis}

\begin{figure}[t]
    \includegraphics[width=87mm]{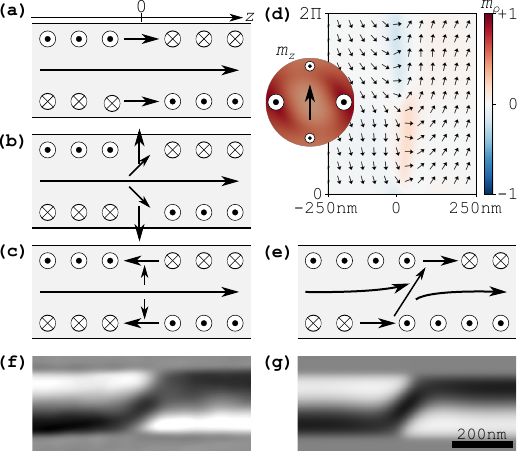}
    \caption{Schematics for DWs. (a)~Néel DW parallel to the axis, (b)~Bloch DW (c)~Néel DW antiparallel to the axis (d)~Unrolled  surface map\refereeAdd{{ }of the $(z,\varphi)$ plane}, and cross\refereeReplace{ }{-}section with sketches for the direction of magnetization, of a simulated DW~(feeLLGood). The color codes the radial and longitudinal components of magnetization\refereeAdd{{ }in these}, respectively.  (e)~Schematics for~(d). (f)~Example of an experimental (ptychography) and (g)~simulated transmission XMCD contrast of a DW of type (d-e).}
    \label{fig-azimuthal-domains-walls}
\end{figure}

\subfigref{fig-azimuthal-domains-walls}{a-c} show\refereeRemove{s} schematics of the simplest rotationally-invariant static DWs that one may \refereeReplace{consider}{anticipate} for nanowires with azimuthal domains. \subfigref{fig-azimuthal-domains-walls}{a,b} depict a Néel and a Bloch DW, respectively\refereeAdd{, as would be named considering the unrolled surface map as a thin film}. In the former case the Néel component is parallel to magnetization on the axis, while \subfigref{fig-azimuthal-domains-walls}c depicts a Néel wall antiparallel to magnetization on the axis. Note that the topology of the three walls is equivalent. In analogy with thin films, from a magnetostatic point of view Néel~(resp. Bloch) walls could be expected to occur for small~(resp. large) radius. However, another specific ingredient needs to be considered, namely, exchange coupling with magnetization on the axis. This adds an extra cost of exchange for the Bloch wall, compared to the Néel wall. Similarly, exchange is expected to lift the degeneracy between the two orientations of the Néel wall, due to the need to create a \qty{180}{\degree} twist with the axis for one orientation of them\bracketsubfigref{fig-azimuthal-domains-walls}{c}. Note that the states sketched in \subfigref{fig-azimuthal-domains-walls}{a-c} are \refereeAdd{also} those expected to occur sequentially upon precession of the DW core around an \OErsted field, such as \refereeReplace{w}{c}ould \refereeAdd{be expected to} occur above a Walker breakdown. Therefore, precession \refereeAdd{would{ }}come\refereeRemove{s} at the expense of a continuously-increasing exchange field, reminiscent of an exchange-spring effect\cite{bib-COE1988,bib-KNE1991,bib-SKO1993}. This is specific to the 3D rotationally-invariant situation, whose topology does not allow to unwind, unlike in the case of a thin film. A crude estimate for this radial exchange energy for the Néel wall in \subfigref{fig-azimuthal-domains-walls}{c}, per unit of wall length along the wire periphery\refereeAdd{, is}:
\begin{equation}
\mathcal{E}_\mathrm{ex}=A\left({\frac{\pi}{R}}\right)^2 R\delta\sim A\pi^2\;,
\label{eq:exchange}
\end{equation}
where $A$ is the micromagnetic exchange stiffness, $\delta$ is the wall width, to first order assumed similar to the radius~$R$\cite{bib-FRU2015b}. Let us make the analogy with the DMI lifting the degeneracy between the two directions of a Néel~DW:
\begin{equation}
\mathcal{E}_\mathrm{DMI}\approx\pi D_\mathrm{V}R/2\;,
\label{eq:dmi}
\end{equation}
with $D_\mathrm{V}$ the volume DMI coefficient. The comparison between \eqnref{eq:exchange} and \ref{eq:dmi} yields:
\begin{equation}
D_\mathrm{V}\approx 2A\pi/R\;.
\label{eq:dmi-value}
\end{equation}
Applied to the typical values $A=\qty{1E-11}{\joule\per\meter}$ and \refereeRemove{radius }$R=\qty{100}{\nano\meter}$, one gets $D_\mathrm{V}=\qty{0.6}{\joule\per\square\meter}$, which is comparable to the order of magnitude of the  interfacial DMI measured in some ultrathin films\cite{bib-KUE2023}. Such an effective value of DMI is expected to enhance the Walker threshold and the maximum DW speed by up to one order of magnitude, under either magnetic field or spin-polarized current\cite{bib-THI2012}. Besides, this simple analogy considers a twist of half a turn only, and neglects the magnetostatic energy arising from the twist, which would further stabilize the DW.

We conducted micromagnetic simulations to evaluate the situation more in depth. We used both feeLLGood\cite{bib-FEE} and MuMax\cite{bib-VAN2014} software, yielding similar results~(see End Matter). \subfigref{fig-azimuthal-domains-walls}{d} shows the surface map and cross-section of a \refereeReplace{domain-wall}{simulated DW} at rest. It is similar to the Néel DW sketched in \subfigref{fig-azimuthal-domains-walls}{a}, with two differences: first, there is a partial Bloch-type component of magnetization, \ie, transverse to the wire, \refereeRemove{and }radial and antiparallel on opposite sides of the wire; second, its intercepts with the wire surface are axially-shifted one with the other. This arrangement is similar to so-called asymmetric Néel walls in films with in-plane magnetization and intermediate thickness, with a mixed Bloch/Néel core and two Néel caps of the same orientation\cite{bib-LAB1969,bib-HUB1969}\bracketsubfigref{fig-azimuthal-domains-walls}{e}. We believe that this arrangement, breaking the rotational symmetry, allows to reduce the magnetostatic energy thanks to the Bloch component largely avoiding volume charges, the Néel caps avoiding surface charges, and globally minimizing exchange energy with the axis, as compared to the case in \subfigref{fig-azimuthal-domains-walls}{b}. \figref{fig-azimuthal-domains-walls}{f,g} show an experimental and simulated XMCD contrast of the DW, the latter computed from the simulated domain wall at rest\cite{bib-FRU2015c}, with blurring adjusted to match the experimental spatial resolution considered at the wire edges\refereeAdd{, and azimuthal angle to provide the best match}. The similarity between both suggests that the simulations describe\refereeRemove{s} accurately the experimental system.

\begin{figure}[t]
    \includegraphics[width=87mm]{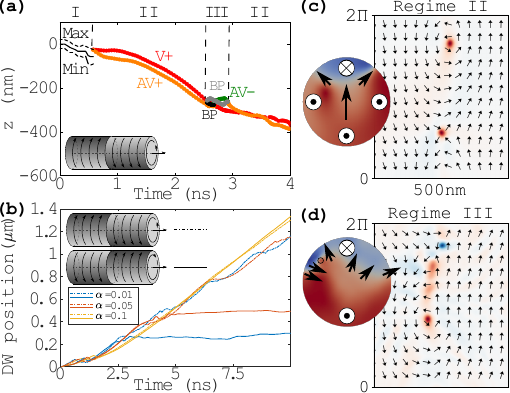}
    \caption{Simulated DW motion. (a)~Longitudinal position of DW versus time for $j=\qty{1E11}{\joule\per\meter\squared}$ and $\alpha=0.01$~(feeLLGood), with three regimes identified as I, II and III, and domains positive with respect to the DW propagation direction (inset). The graph displays the position of Min, average and Max of the radial component at the periphery for regime~I, surface vortex~(V)/antivortex~(AV) along with their polarity in regime~II, with addition Bloch points in regime~III (b)~DW displacement versus time for three values of damping and the two cases of circulation of domains on either side of the DW~(MuMax) (c,d)~Unrolled surface maps and cross\refereeReplace{ }{-}sections for $j=\qty{1E11}{\joule\per\meter\squared}$ and $\alpha=0.01$, typical of regimes II and III from (a), at $\qty{1.841}{\nano\second}$ and $\qty{2.842}{\nano\second}$, respectively, while regime~I is qualitatively described by the static distribution shown in \subfigref{fig-azimuthal-domains-walls}d with the same color scale~(feeLLGood).}
    \label{fig-simulations-motion}
\end{figure}

We \refereeAdd{then} simulated the response of the DW to an applied current, considering the \OErsted field as the sole driving force of motion\cite{bib-FRU2021c}, and covered damping coefficients in the range $0.01$ to~$0.1$. \refereeAdd{Spin-transfer torques are neglected, because the values of current considered around $\qty{E11}{\ampere\per\squared\meter}$ are typically one order of magnitude smaller than the expected current-driven Walker threshold\cite{bib-THI2008}, while the resulting \OE rsted field is $\qty{6.3}{\milli\tesla}$ at the periphery of a $\qty{200}{\nano\meter}$-diameter, which is several-fold the value of Walker field\cite{bib-THI2006}. We checked that considering spin-transfer torque in addition to \OE rsted field only very marginally changes the behavior. This contrasts with the situation of tubes, in which the ratio of field versus current torques depends on the thickness of the tube wall\cite{bib-FRU2021c}.} For large damping and low current density the motion is steady-state, with speed \qty{100}{\meter\per\second} or lower, which we will call regime~I. Surface maps reveal only quantitative changes compared with \subfigref{fig-azimuthal-domains-walls}d~(not shown here). For lower damping or larger current density, one or even two successive \refereeReplace{disruptions}{qualitative changes of the internal structure} of \refereeAdd{the} DW \refereeRemove{motion} occur over time, defining in total three regimes, which we labeled I to~III on \subfigref{fig-simulations-motion}a. \refereeReplace{Each disruption is associated with a qualitative transformation}{The structure} of the DW\refereeReplace{, as}{ in regimes II and III is} illustrated by their surface map and sketch of interior on \subfigref{fig-simulations-motion}{c,d}. \refereeRemove{It is consistent with the regime of DW motion below the Walker breakdown in the 1d model, \ie, speed increases with applied current and decreases with damping.}\refereeReplace{T}{At t}he transition \refereeAdd{from regime~I} to regime~II \refereeRemove{occurs when the speed reaches approximately $\qty{200}{\meter\per\second}$.}\refereeReplace{M}{m}agnetization becomes fully radial at a given \refereeAdd{peripheral }location, \refereeReplace{leading to}{followed by} the creation of a pair of surface vortex/antivortex\refereeRemove{ pair} that \refereeReplace{then}{later on} move apart azimuthally. This dynamical process is \refereeAdd{topologically} analogous to others occurring for DWs in cylindrical nanowires \refereeAdd{with axial domains} under both applied field\cite{bib-FRU2019} or current\cite{bib-FRU2021,bib-FRU2025}. \refereeReplace{M}{The longitudinal m}agnetization in the peripheral area between the two surface objects has then rotated by $\pi$ with respect to the DW at rest, \ie, similar to the Néel wall shown in \subfigref{fig-azimuthal-domains-walls}c. For intermediate values of current and damping the regime~II reaches a steady state, invariant \refereeAdd{over time{ }}upon a combined translation/rotation. \refereeRemove{In this case, the process looks like the beginning of the precession of the internal degree of freedom of the DW, however reaching a dynamical equilibrium angle of $\ang{180}$ and not developing a full Walker process. }For still lower damping or larger current density, a transition to regime~III occurs \refereeReplace{for}{when the} DW speed \refereeAdd{reaches }around \qty{200}{\meter\per\second}, through the nucleation of a pair of Bloch and anti-Bloch points in the bulk of the wire\bracketsubfigref{fig-azimuthal-domains-walls}a. Shortly afterwards, the latter annihilates at the surface\refereeReplace{ and}{,} thereby changing the polarity of the surface antivortex to negative~(see Ref.\citenum{bib-FRU2025} for the conservation of the total topological charge). This process is analogous \refereeReplace{with}{to} the switching of vortex core \refereeAdd{in flat disks }during gyration, occurring around the same drift speed\cite{bib-VAN2006,bib-HER2007,bib-YAM2007}. \refereeReplace{This process}{It} comes with a sharp drop of DW mobility\refereeRemove{, reminiscent of the Walker process in the 1d model\cite{bib-NAK2005}}. However, \refereeRemove{in turn and }soon afterwards, the second Bloch point \refereeAdd{in turn }reaches the surface and switches the antivortex back to its initial positive polarity. The regime \refereeReplace{is again}{turns back to} II, and the speed of the DW increases again. \refereeRemove{Thus, a Walker-like process with a $\ang{360}$ rotation of magnetization occurs in regime III but does not affect the entire DW, remaining close to the wire periphery and soon collapsing. }Later on the DW transitions regularly and temporarily to regime~III, with a frequency increasing with the magnitude of the current. \refereeReplace{We believe that the dynamical equilibrium of regime~II and the prevention of a full Walker process in regime III result from an exchange-spring effect with the axial core, preventing the process to reach the axis of the wire. Thus, o}{O}verall\refereeAdd{,} the DW speed remains high.

\refereeAdd{The above illustrated motion for a given value of damping coefficient, and for a DW separating domains $C-$ and $C+$ on the left- and right-hand side, respectively. We checked that the results are exactly the same for both positive or negative current, \ie, for the direction of DW propagation parallel to the axial vector of the domain circulation towards which the DW is moving. To broaden this picture, }\subfigref{fig-simulations-motion}b illustrates DW motion over time \refereeAdd{for different values of damping coefficients, and DWs separating domains from either $C-$ to $C+$ or from $C+$ to $C-$}. \refereeRemove{Note that all of the above has been simulated for DW propagation direction parallel to the axial vector for domain circulation on either side, be it for positive or negative \OErsted field: $C+$ on the right-hand side and $C-$ on the left-hand side~(dotted line in \subfigref{fig-simulations-motion}b).} While regimes I and II occur similarly for DW propagation now antiparallel to the axial vector for domain circulation on either side, there is a drastic difference for regime III, which remains indefinite with a very low speed~(dotted line in \subfigref{fig-simulations-motion}b). We believe that this dramatic non-reciproc\refereeReplace{ity}{al} effect is induced by curvature, analogous to that of spin-wave propagation\cite{bib-OTA2016,bib-OTA2017}. \refereeReplace{B}{Indeed, b}oth spin-wave and DW motion \refereeRemove{indeed} share the Damon-Eshbach geometry\cite{bib-STA2009} with propagation direction and peripheral magnetization $\vect M$ perpendicular. Non-reciprocal effects of DW motion due to curvature have been reported experimentally already, \eg, in twisted ribbons\cite{bib-FAR2025}. However, note that the time for the first transition to regime III is a few nanoseconds, \ie, longer than the experimental pulse length. We therefore expect that regime III may not be relevant in the present experiments, explaining the absence of non-reciprocity\refereeAdd{{ }in the experimental results}\bracketsubfigref{fig-current-pulses}b. \refereeRemove{Nevertheless, the apparent experimental speed is 2 to 3 times larger than the simulated one. Further simulations is needed to fully explain this discrepancy, provide more details about the different dynamic regimes, and the microscopic understanding of the non-reciprocal effect probably deserve more extensive simulations. Intrinsic (\eg, strength of spatial profile of anisotropy) or extrinsic (\eg, granular structure) effects may also contribute to further delaying the Walker breakdown.}

\refereeAdd{
We now discuss these results, with a view to highlight both similarities and differences with DW motion as known in lower dimension, either in the 1D model or in 2D flat strips\cite{bib-THI2006}. We restrict the discussion to field-driven motion, as the \OE rsted field is by far the main stimulus in the present situation. In the 1D model, the DW internal degree of freedom rotates with respect to its equilibrium direction but reaches a limiting value $<\pi/4$, leading to steady-state motion of the DW. The DW moves largely due to the torques induced by its internal field of either dipolar\refereeReplace{ of}{,} anisotropy \refereeAdd{{or DMI }}origin, with speed increasing with applied field and decreasing with damping. Above a threshold of field, the process is qualitatively different: the DW internal degree of freedom rotates unevenly but periodically by $2\pi$, with speed oscillating accordingly. The mean value of speed still increases with applied field but is now inversely proportional to damping, therefore it is much lower than in the steady-state regime. Due to the abrupt change of regime and the sharp drop of velocity, the threshold process is called the Walker breakdown\cite{bib-SCH1974}. In strips, \ie, a 2D system, the picture in terms of speed is the same. Nevertheless, while below the Walker breakdown the DW is mostly invariant across the strip and similar to the 1D model, above the Walker breakdown the rotation of the internal degree of freedom of the DW is incoherent along the DW length, taking the form of the nucleation-propagation of vortices or antivortices nucleated at the strip edge. A generalized DW internal angle is sometimes defined, reflecting the transverse location of the vortex/antivortex, and whose value oscillates periodically and continuously such as in the 1D model. In the 3D present situation, the phenomenology of regime I is consistent with DW motion below the Walker breakdown in the 1D model, \ie, steady-state motion and speed increasing with applied current and decreasing with damping. Regime~II shares with the 2D situation the nucleation of a pair of vortex/antivortex at the wire surface, and their moving one apart from the other across the DW peripheral length, \ie, with an increase of the generalized internal angle. However, in contrast to 2D strips, their peripheral motion reaches a steady-state angle, \ie, the DW reaches a steady-state form, for which the DW peripheral magnetization has rotated by $\pi$ only between the surface vortex/antivortex\bracketsubfigref{fig-simulations-motion}c. The initial nucleation of a pair of vortex/antivortex is consistent with the 2D situation above the Walker breakdown, however the steady-state motion is consistent with the 2D situation below the Walker breakdown, along with an increase of speed with applied field and decreases with damping. This leads us to consider that the Walker process has been initiated but is hampered by an exchange-spring effect with the core of the wire, which is axially-magnetized along a given direction. Note that this physical effect is different from the one already identified in homogeneously axially-magnetized nanowires and nanotubes, related to magnetostatics\cite{bib-YAN2011b,bib-HER2016,bib-FRU2019b,bib-BRA2023}. This effect is analogous to a situation considered in strips, namely a DW moving through a comb of uniformly-magnetized transverse branches playing a rectifying role on the DWs induced by exchange coupling with the transverse strips of the comb\cite{bib-LEW2010}. Nevertheless, a difference with the 2D case is the continuous azimuthal rotation of the DW and its surface vortex/antivortex pair during its steady-state forward motion. This feature is made possible by the extra azimuthal degree of freedom existing in the 3D case, and the geometric rotational invariance, compared to the 1D and 2D situations. We now turn to discussing regime~III. It is interesting to notice the quantitative similarity with vortices moving in a 2D system, whose core switches of polarity when its speed exceed $\approx\qty{200}{\meter\per\second}$, \ie, similar to the DW speed at the end of regime~II. However, while in the 2D case the associated Bloch points are quickly expelled from the thin film, in the present 3D situation Bloch points have the option to remain in the bulk of the system. For DWs moving towards a domain with antiparallel axial vector of its circulation~(continuous lines in \subfigref{fig-simulations-motion}b), the Bloch point indeed remains dynamically below the surface. Motion suffers an abrupt breakdown of DW speed, the latter increasing with damping, which is phenomenologically similar to an above-Walker situation. However, motion is steady state, unlike the 1D/2D Walker process. This hybrid situation has therefore no equivalent in lower dimension. Conversely, for DWs moving towards a domain with parallel axial vector of its circulation~(dotted lines in \subfigref{fig-simulations-motion}b), the periodic expulsion of Bloch point and switching back to regime~II allows to sustain a large speed, which is largely independent of the damping parameter. Note that the non-reciprocity depends on the sequence of domain circulation on either side of the DW, \ie, left/right or right/left, but for a given sequence it dose not depend on the direction of motion. This is because on either side of a DW both the direction of motion and the circulation switch, so that the situation is similar from the perspective of the Damon-Eshbach geometry.

The experiments and simulations conducted, presented and analyzed above, open new perspectives and call for further investigation and understanding. The apparent experimental speed is indeed 2 to 3 times larger than the highest simulated one, and we can also not tell for sure to which regime it belongs with. We identify several directions to address this. First, on the material side, there is a need to design wires with a sufficiently-wide operating window of charge current to correlate experiments with the different regimes expected\bracketsubfigref{fig-simulations-motion}b, lowering the pinning current by reducing the microstructure~(lower bound)\futurecite{correlative microscopy} and increasing the current threshold inducing coherent switching, by increasing the strength of azimuthal anisotropy~(upper bound). Second, on the measurement side and for the same sake, there is a need to measure DW motion under the stimulus of pulses longer than a few nanoseconds, to possibly access the transition to regime~III, to seek the confirmation of both the breakdown of speed and its dramatic non-reciprocity. Finally, further simulations and theoretical analysis may bring additional light on the microscopic mechanisms underlying non-reciprocity, as well as the possible role of intrinsic (\eg, strength of spatial profile of anisotropy) and extrinsic (\eg, granular structure) effects in further delaying the Walker breakdown.%
}

\section{Conclusion}

We have measured unexpectedly-fast \OErsted-field-driven domain wall~(DW) motion in cylindrical nanowires with azimuthal peripheral magnetization~($\approx\qty{600}{\meter\per\second}$). Micromagnetic simulations reveal that several domain-wall transformations occur up to the onset of a Walker process, involving surface vortices/antivortices and Bloch points in the sub-surface area. However this process remains localized at the wire periphery and \refereeReplace{soon collapses, }{may }preserv\refereeReplace{ing}{e} the high DW mobility\refereeReplace{, and}{. It} is also \refereeAdd{dramatically} non-reciprocal depending on the set of \refereeRemove{two }domain circulations on either side of a given DW\refereeAdd{, giving rise or not to a sharp collapse of DW speed}. Both effects are specific to three-dimensional nanomagnetism, in which a topology-locked exchange-spring effect between the peripheral and the core magnetization prevents the Walker breakdown, and curvature induces non-reciprocity. \refereeReplace{As such, it is}{Both effects are} expected to be specific to nanowires with radius of the order of $\qty{100}{\nano\meter}$, for which exchange energy is sizable\refereeAdd{, but still allows for azimuthal magnetization}.

\section{Acknowledgments}

We thank L.~Vila, W.~Savero Torres and K.~Garello for assistance with the magnetoresistance measurements. We acknowledge support from the team of the Nanofab platform (Institut Néel) and the French RENATECH network implemented at the Upstream Technological Platform in Grenoble PTA  (ANR-22-PEEL-0015). The work was partially supported by Projects PID2023-150853NB-C31, PID2024-155385NB-C31 and PID2024-155385NA-C32, funded by MICIU/AEI /10.13039/501100011033 and by FEDER, UE and by Comunidad de Madrid via grant TEC-2024/TEC-380 (Mag4TIC-CM), and the QuantAlps research federation FR2053 in Grenoble. A CC-BY public copyright license has been applied by the authors to the present document and will be applied to all subsequent versions up to the Author Accepted Manuscript arising from this submission, in accordance with the grant’s open access conditions \cite{bib-CC-BY}.

\section{End matter}

\subsection{Fabrication}

Permalloy cylindrical nanowires of \qty{200}{\nano\meter} diameter with compositional modulations along their length were synthesized by template-assisted electrochemical deposition. Nanoporous anodic aluminium oxide (AAO) templates were prepared by one-step hard anodization of Al disks with the pores opened to a final diameter of \qty{200}{\nano\meter}\cite{bib-DOM2021}. Electrodeposition was carried out under conditions similar to those described in \cite{bib-RUI2018}, using an electrolyte composed of NiSO$_4$ (\qty{0.47}{M}), NiCl$_2$ (\qty{0.01}{M}), FeSO$_4$ (\qty{0.09}{M}) and H$_3$BO$_3$ (\qty{0.4}{M}). Applied voltage pulses of \qty{-1.35}{\volt} during \qty{200}{\second} led to the deposition of permalloy (Fe$_{20}$Ni$_{80}$) segments with a length of \qty{3.6}{\micro\meter}, while short pulses of \qty{-1.1}{\volt} for \qty{20}{\second} allowed the deposition of chemical modulations with a composition of Fe$_{70}$Ni$_{30}$, \qtyrange{40}{60}{nm} in length.

After growth, the nanowires were released from the template and dispersed on a Si substrate with \qty{40}{\nano\meter}-thick Si$_3$N$_4$ windows to allow X-ray transmission for TXM and STXM/ptychography experiments. Two \ce{Au/Ti} contact pads were made in each nanowire with laser lithography. The Si chip was then mounted on a dedicated printed cardboard~(PCB) equipped with SMP connectors, which were connected with a flexible cable to an SMP/SMA port of the microscope vacuum vessel. The two electric pads were connected to the PCB with microbonding. Based on the two-point resistances measured, the geometry of the wires (diameter and distance between the two contacts) and neglecting the resistance of the leads, the resistivity of the wires was $\rho$ = $\qty{2.8E-7}{\ohm\meter}$ with a low dispersion from wire to wire. This is similar to the bulk value, around $\rho$ = $\qty{2E-7}{\ohm\meter}$ at room temperature\cite{bib-COU1996}, and lower than permalloy in physical vapor deposition thin films, around \qtyrange{5E-7}{6E-7}{\ohm\meter} for films \qty{100}{\nano\meter} and thicker\cite{bib-LO1987,bib-COU2006}.

\subsection{Transmission X-ray Microscopy and Ptychography}
XMCD measurements were performed either at the MISTRAL beamline in ALBA synchrotron using a transmission X-ray microscope (TXM) in static conditions, or at the HERMES beamline in SOLEIL synchrotron using ptychography, in static or time-resolved conditions.

TXM at MISTRAL uses a glass capillary to focus the beam on the sample field of view, and a zone plate to construct the image and record it with a camera. TXM has a spatial resolution of about \qty{25}{\nano\meter}. The photon energy was tuned to the Fe L$_3$ absorption edge.

Pytchography is an extension of scanning transmission X-ray microscopy (STXM), recording far-field diffraction patterns for overlapping 2D rastered-scanned points on the sample, with a probe size $\approx\qty{50}{\nano\meter}$. An iterative algorithm allows to reconstruct both the intensity and the phase of X-rays at every location of the sample, with a spatial resolution around \qty{10}{\nano\meter}. When combined with XMCD, ptychography provides information about the component of magnetization parallel to the X-ray beam. We set the X-ray energy around \qty{1}{eV} below the Fe L$_3$ edge, which decreases absorption while maintaining a significant magnetic scattering, suitable to image thick samples as in the present case\cite{bib-NEE2024}. To obtain time resolution, an oscillatory ac current was fed inside the wire at the cavity radio frequency of the ring, \qty{352.202}{MHz}, and synchronized with the clock signal from the synchrotron. Consequently, in the case of reproducible magnetic processes, every X-ray bunch probes the same distribution of magnetization. Adjusting the time delay between pump and probe over one period allows one to build a magnetic video of the magnetization process. More details of the technique will be published elsewhere.

\subsection{Electric current pulses}
In this section, we present the method used to determine the duration of the experimental current pulses and the role of Joule heating.
\refereeAdd{We use a Kentech UTV50p pulse generator, which generates $\qty{50}{\ohm}$-adapted voltage pulses up to $\qty{50}{\volt}$ and slightly sub-ns rise time.} \refereeAdd{We use a Siglent} \refereeReplace{$\qty{100}{MHz}$}{$\qty{2}{GHz}$} oscilloscope to monitor two pulses of current: first, after the pulse generator and before sample, using a \si{\mega\ohm} load, and second, after the sample, using a \qty{50}{\ohm} load. The latter allows to determine the current flowing through the sample, while a significant part of the pulse is reflected due to the resistance mismatch with the sample. We checked that the sample resistance inferred either from a dc measurement with a multimeter, or from the height of pulses sent, reflected and transmitted, match together.

\begin{figure}[h]
    \centering
    \includegraphics[width=1\linewidth]{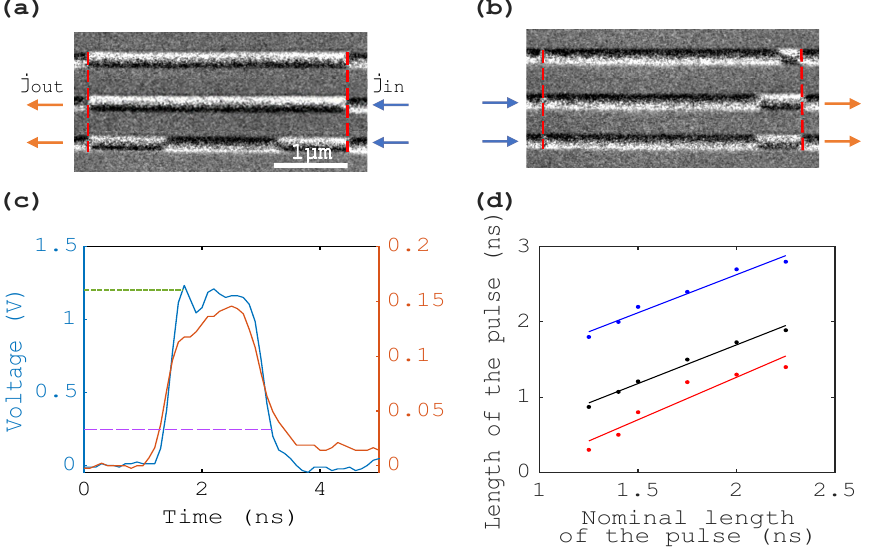}
    \caption{(a,b) The top TXM images show initial states. The following images in each sequence are after applying current pulses to the initial state of (a) $\qty{1.04E11}{}$ and $\qty{1.1E11}{\ampere\per\meter\squared}$  and (b) $\qty{2.2E10}{}$  and $\qty{6.9E9}{\ampere\per\meter\squared}$, respectively. (c) Injected (blue) and transmitted (orange) pulse shape for a pulse with nominal voltage $\qty{1.2}{\volt}$. The green and purple dashed lines correspond to the minimum injected voltage of the incoming pulse required for DW nucleation and motion, respectively. (d) Upper (blue) and lower (red) bounds, and the effective experimental (black) length of the pulse measured after the sample, versus the nominal length, and their corresponding linear fits. \dataref{ Sample R10 measured in ALBA, Nov2023}}
    \label{pulselength}
\end{figure}

The pulse length was determined from the injected current pulses using current-density thresholds. Since the data in \figref{fig-current-pulses} is based on the initial nucleation of a domain, we define that the pulse begins when the measured voltage corresponds to the minimum current density required for domain nucleation $\qty{1.1E11}{\ampere\per\meter\squared}$. Similarly, we define that the pulse ends at the minimum current density allowing DW motion $\qty{2.51E10}{\ampere\per\meter\squared}$. These two values were obtained from the experimental results shown in Figure \ref{pulselength}(a,b). To  determine the error bars, the upper bound of the pulse length was considered to be the total length of the measured pulse, while the lower bound corresponds to the length of the plateau. The upper and lower bounds and the effective pulse length are represented in Figure \ref{pulselength}(c,d).

An upper bound of the increase of temperature $\Delta$T due to Joule heating in a nanowire can be calculated as:
    \begin{equation}
        \Delta T = \frac{j^2\rho t}{Dc}\;
    \end{equation}
where j and \textit{t} are the current density and duration of the electric pulses and $\rho$, \textit{D}, \textit{c} are the resistivity, density, and the specific heat capacity per unit mass of the nanowire, respectively. For the case of a 240 nm-diameter permalloy nanowire, we use our experimentally measured value of $\rho$ = $\qty{2.8E-7}{\ohm\meter}$ and literature values for  $c=\qty{1.1}{\joule\per\gram\per\kelvin}$ for Ni nanowires, and bulk density $D=\qty{8.6}{\gram\per\meter\cubed}$. For the maximum length of the current pulse, $\qty{2.25}{\nano\second}$,  and a current density of $j=\qty{1.1E11}{\ampere\per\meter\squared}$, we estimate a maximum increase of temperature of $\qty{0.8}{\kelvin}$. Hence, we consider Joule heating to be negligible in the study of domain wall motion under current pulses in this system. This arises thanks to the low current density, good electrical conductivity of the material, short pulse
length and single-pulse usage.

\subsection{Coherent switching for large current density}

\begin{figure}[h]
    \centering
    \includegraphics[width=0.9\linewidth]{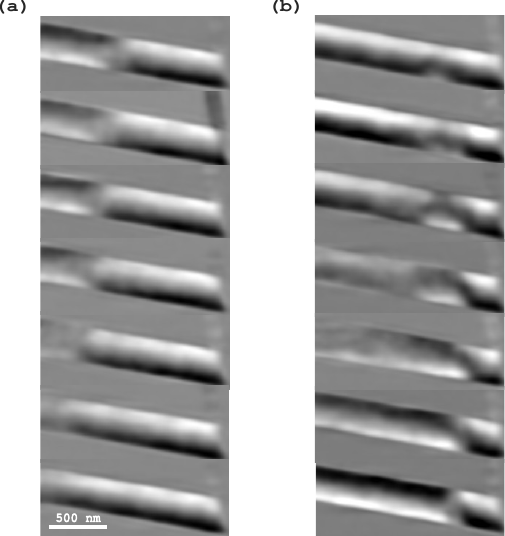}
    \caption{Time-resolved ptychography images. The nanowire was excited with an AC current at \qty{352.202}{\mega\hertz} and a current density of (a) $\qty{3E10}{\ampere\per\square\meter}$ and (b) $\qty{4.9E10}{\ampere\per\meter\squared}$. Frames were acquired every \qty{240}{\pico\second}. \dataref{Sample Velvet V8, measured in SOLEIL Oct2024.}}
    \label{TRptycho}
\end{figure}

Converting the DW position measured between two current pulses into DW speed, requires that the underlying mechanism is indeed DW motion. A sanity check is that the inferred speed is largely independent of the pulse length, which is the protocol used in the manuscript. Using time-resolved ptychography imaging, we checked that DW motion is indeed the mechanism at low-enough current density, while coherent reversal can occur at large current\bracketsubfigref{TRptycho}, which can lead to unrealistically-large values of DW speed if interpreted as DW motion. In this experiment, a permalloy nanowire with similar characteristics was investigated under low- and high-current AC excitation. We  believe that the difference of threshold for DW motion and coherent reversal, compared to the wires images with TXM, result from difference in the magnitude of magnetic anisotropy.

\subsection{Evaluation of the value of magnetic anisotropy}

In the micromagnetic simulations, the azimuthal state was promoted by considering a hard-axis contribution to magnetic anisotropy $\Ku$. We evaluated the strength of $\Ku$ by comparing anisotropic magnetoresistance (AMR) measurements of an electrically-contacted individual nanowire, with TetraX micromagnetic simulations.

\begin{figure}[h]
    \centering
    \includegraphics[width=1\linewidth]{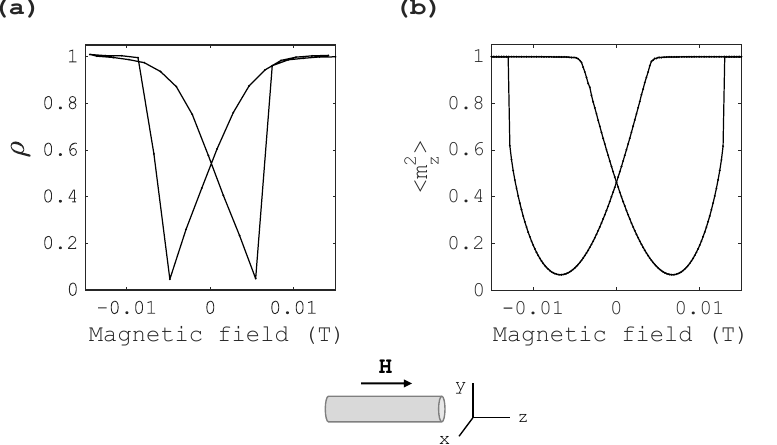}
    \caption{(a) Normalized experimental AMR curve of a \qty{240}{\nano\meter} permalloy nanowire with magnetic field parallel to the axis. (b) Simulated $\langle\mathrm{m_z}^2\rangle$ versus magnetic field applied parallel to the axis for $\Ku=\qty{6E3}{\joule\per\meter\cubed}$. \dataref{Sample F9, measured in ALBA in Dec2024. AMR measurements performed in SPINTEC (Laurent Villa set up) in April 2025.}}
    \label{AMR}
\end{figure}

AMR was measured while applying a magnetic field successively perpendicular and parallel to the wire axis, up to a magnitude sufficient to reach a saturation state. Resistance is larger for magnetization parallel with the current, as usual for Permalloy. By comparing the value of resistances for saturation along the two directions, we determined the magnitude of AMR to be $\Delta_\mathrm{AMR}\approx\qty{3}{\%}$. Note that in the rare cases that a contact resistance adds up, the total resistance is increased and the apparent AMR is decreased. From the apparent AMR, we defined the normalized resistance $\rho=1-(R_\parallel-R)/(R_\parallel\cdot\Delta_{\mathrm{AMR}})$, which we show versus magnetic field applied parallel to the nanowire axis on \subfigref{AMR}{a}. The high resistance level reached under a few tens of \si{\milli\tesla} is associated to magnetization saturation along the axial direction, while the lower values of resistance around remanence result from the curling of magnetization at the periphery of the wire. Similar to the micromagnetic simulations reported in the main text, we considered a \qty{200}{\nano\meter}-diameter permalloy disk, with $\Ms=\qty{800E3}{\ampere\per\meter}$ and $A_\mathrm{ex}=\qty{13E-12}{\joule\per\meter}$. The resulting AMR was computed from the cross-sectional average $\langle\mathrm{m_z}^2\rangle$. The magnitude of the uniaxial anisotropy $\Ku$ was varied systematically. The comparison between the experimental and simulated AMR curves is not straightforward due to the hysteresis, which is larger in the micromagnetic simulation, considering an infinite and translational-invariant wire. In the experimental case, the hysteresis associated with the switching of circulation can be promoted the wire ends, chemical modulations and the existence of domains. We therefore focused the comparison on the first part of the loops, from saturation to remanence and only slightly beyond. By comparing different experimental measurements with the set of simulated curves, we estimate an anisotropy in the range $\Ku=\qtyrange{6E3}{1E4}{\joule\per\meter\cubed}$. We subsequently used the latter in the micromagnetic simulations presented in the manuscript, as it better fits the details of the magnetic configuration of the domain wall\bracketsubfigref{fig-azimuthal-domains-walls}{f-g}.

\subsection{Micromagnetic simulations}

Material parameters are identical in feeLLGood and MuMax: magnetization $\Ms=\qty{800}{\kilo\ampere\per\meter}$, exchange stiffness $A_\mathrm{ex}=\qty{13E12}{\joule\per\meter}$, hard-axis uniaxial anisotropy $\Ku=\qty{1E4}{\joule\per\cubic\meter}$, gyromagnetic ratio $\gamma=\qty{1.7595E11}{\tesla\per\meter}$. The volume of the cells is \qty{8}{\nano\meter\cubed} cubic cells for MuMax, \qty{8.6}{\nano\meter\cubed} tetrahedra on the average for feeLLGood, meshed with Gmsh\cite{bib-GEU2009} using an extrusion method along the axis of the wire.

The time-dependent position of the DW in MumaX simulations was determined from the spatial distribution of the longitudinal component of magnetization $m_z$ at the periphery of the wire. For each time step, we consider two cross-sections parallel to the wire axis, one near the upper and one near the lower edges of the nanowire. Along each, we determined the minimum value of the azimuthal component, which identifies the position of the DW. Because the DW is no rotationnally-invariant, these two values may be slightly different. To minimize this effect, the DW position was computed as the average of these two values.


\begin{thebibliography}{58}%
\makeatletter
\providecommand \@ifxundefined [1]{%
 \@ifx{#1\undefined}
}%
\providecommand \@ifnum [1]{%
 \ifnum #1\expandafter \@firstoftwo
 \else \expandafter \@secondoftwo
 \fi
}%
\providecommand \@ifx [1]{%
 \ifx #1\expandafter \@firstoftwo
 \else \expandafter \@secondoftwo
 \fi
}%
\providecommand \natexlab [1]{#1}%
\providecommand \enquote  [1]{``#1''}%
\providecommand \bibnamefont  [1]{#1}%
\providecommand \bibfnamefont [1]{#1}%
\providecommand \citenamefont [1]{#1}%
\providecommand \href@noop [0]{\@secondoftwo}%
\providecommand \href [0]{\begingroup \@sanitize@url \@href}%
\providecommand \@href[1]{\@@startlink{#1}\@@href}%
\providecommand \@@href[1]{\endgroup#1\@@endlink}%
\providecommand \@sanitize@url [0]{\catcode `\\12\catcode `\$12\catcode
  `\&12\catcode `\#12\catcode `\^12\catcode `\_12\catcode `\%12\relax}%
\providecommand \@@startlink[1]{}%
\providecommand \@@endlink[0]{}%
\providecommand \url  [0]{\begingroup\@sanitize@url \@url }%
\providecommand \@url [1]{\endgroup\@href {#1}{\urlprefix }}%
\providecommand \urlprefix  [0]{URL }%
\providecommand \Eprint [0]{\href }%
\providecommand \doibase [0]{https://doi.org/}%
\providecommand \selectlanguage [0]{\@gobble}%
\providecommand \bibinfo  [0]{\@secondoftwo}%
\providecommand \bibfield  [0]{\@secondoftwo}%
\providecommand \translation [1]{[#1]}%
\providecommand \BibitemOpen [0]{}%
\providecommand \bibitemStop [0]{}%
\providecommand \bibitemNoStop [0]{.\EOS\space}%
\providecommand \EOS [0]{\spacefactor3000\relax}%
\providecommand \BibitemShut  [1]{\csname bibitem#1\endcsname}%
\let\auto@bib@innerbib\@empty
\bibitem [{\citenamefont {Thiaville}\ and\ \citenamefont
  {Nakatani}(2006)}]{bib-THI2006}%
  \BibitemOpen
  \bibfield  {author} {\bibinfo {author} {\bibfnamefont {A.}~\bibnamefont
  {Thiaville}}\ and\ \bibinfo {author} {\bibfnamefont {Y.}~\bibnamefont
  {Nakatani}},\ }\bibinfo {title} {Spin dynamics in confined magnetic
  structures {III}}\ (\bibinfo  {publisher} {Springer},\ \bibinfo {address}
  {Berlin},\ \bibinfo {year} {2006})\ Chap.\ \bibinfo {chapter} {Domain-wall
  dynamics in nanowires and nanostrips}, pp.\ \bibinfo {pages}
  {161--205}\BibitemShut {NoStop}%
\bibitem [{\citenamefont {Thiaville}\ and\ \citenamefont
  {Nakatani}(2009)}]{bib-THI2008}%
  \BibitemOpen
  \bibfield  {author} {\bibinfo {author} {\bibfnamefont {A.}~\bibnamefont
  {Thiaville}}\ and\ \bibinfo {author} {\bibfnamefont {Y.}~\bibnamefont
  {Nakatani}},\ }\bibinfo {title} {Nanomagnetism and spintronics}\ (\bibinfo
  {publisher} {Elsevier},\ \bibinfo {year} {2009})\ Chap.\ \bibinfo {chapter}
  {Micromagnetic simulation of domain wall dynamics in nanostrips}, pp.\
  \bibinfo {pages} {231--276}\BibitemShut {NoStop}%
\bibitem [{\citenamefont {Thomas}\ and\ \citenamefont
  {Parkin}(2007)}]{bib-THO2007}%
  \BibitemOpen
  \bibfield  {author} {\bibinfo {author} {\bibfnamefont {L.}~\bibnamefont
  {Thomas}}\ and\ \bibinfo {author} {\bibfnamefont {S.~S.~P.}\ \bibnamefont
  {Parkin}},\ }\bibinfo {title} {Micromagnetism}\ (\bibinfo  {publisher}
  {Wiley},\ \bibinfo {address} {Chichester, England},\ \bibinfo {year} {2007})\
  Chap.\ \bibinfo {chapter} {Current induced domain-wall motion in magnetic
  nanowires}, pp.\ \bibinfo {pages} {942--982}\BibitemShut {NoStop}%
\bibitem [{\citenamefont {Mougin}\ \emph {et~al.}(2007)\citenamefont {Mougin},
  \citenamefont {andJ.P. Adam}, \citenamefont {Metaxas},\ and\ \citenamefont
  {Ferré}}]{bib-MOU2007}%
  \BibitemOpen
  \bibfield  {author} {\bibinfo {author} {\bibfnamefont {A.}~\bibnamefont
  {Mougin}}, \bibinfo {author} {\bibfnamefont {M.~C.}\ \bibnamefont {andJ.P.
  Adam}}, \bibinfo {author} {\bibfnamefont {P.}~\bibnamefont {Metaxas}},\ and\
  \bibinfo {author} {\bibfnamefont {J.}~\bibnamefont {Ferré}},\ }\emph
  {\bibinfo {title} {Domain wall mobility, stability and walker breakdown in
  magnetic nanowires}},\ \href {https://doi.org/10.1209/0295-5075/78/57007}
  {\bibfield  {journal} {\bibinfo  {journal} {\epl}\ }\textbf {\bibinfo
  {volume} {78}},\ \bibinfo {pages} {57007} (\bibinfo {year}
  {2007})}\BibitemShut {NoStop}%
\bibitem [{\citenamefont {Beach}\ \emph {et~al.}(2005)\citenamefont {Beach},
  \citenamefont {Nistor}, \citenamefont {Knuston}, \citenamefont {Tsoi},\ and\
  \citenamefont {Erskine}}]{bib-BEA2005}%
  \BibitemOpen
  \bibfield  {author} {\bibinfo {author} {\bibfnamefont {G.~S.~D.}\
  \bibnamefont {Beach}}, \bibinfo {author} {\bibfnamefont {C.}~\bibnamefont
  {Nistor}}, \bibinfo {author} {\bibfnamefont {C.}~\bibnamefont {Knuston}},
  \bibinfo {author} {\bibfnamefont {M.}~\bibnamefont {Tsoi}},\ and\ \bibinfo
  {author} {\bibfnamefont {J.~L.}\ \bibnamefont {Erskine}},\ }\emph {\bibinfo
  {title} {Dynamics of field-driven domain-wall propagation in ferromagnetic
  nanowires}},\ \href {https://doi.org/10.1038/nmat1477} {\bibfield  {journal}
  {\bibinfo  {journal} {\NatMater}\ }\textbf {\bibinfo {volume} {4}},\ \bibinfo
  {pages} {741} (\bibinfo {year} {2005})}\BibitemShut {NoStop}%
\bibitem [{\citenamefont {Hayashi}\ \emph
  {et~al.}(2007{\natexlab{a}})\citenamefont {Hayashi}, \citenamefont {Thomas},
  \citenamefont {Rettner}, \citenamefont {Moriya}, \citenamefont {Bazaliy},\
  and\ \citenamefont {Parkin}}]{bib-HAY2007b}%
  \BibitemOpen
  \bibfield  {author} {\bibinfo {author} {\bibfnamefont {M.}~\bibnamefont
  {Hayashi}}, \bibinfo {author} {\bibfnamefont {L.}~\bibnamefont {Thomas}},
  \bibinfo {author} {\bibfnamefont {C.}~\bibnamefont {Rettner}}, \bibinfo
  {author} {\bibfnamefont {R.}~\bibnamefont {Moriya}}, \bibinfo {author}
  {\bibfnamefont {Y.~B.}\ \bibnamefont {Bazaliy}},\ and\ \bibinfo {author}
  {\bibfnamefont {S.~S.~P.}\ \bibnamefont {Parkin}},\ }\emph {\bibinfo {title}
  {Current driven domain wall velocities exceeding the spin angular momentum
  transfer rate in permalloy nanowires}},\ \href
  {https://doi.org/10.1103/physrevlett.98.037204} {\bibfield  {journal}
  {\bibinfo  {journal} {\prl}\ }\textbf {\bibinfo {volume} {98}},\ \bibinfo
  {pages} {037204} (\bibinfo {year} {2007}{\natexlab{a}})}\BibitemShut
  {NoStop}%
\bibitem [{\citenamefont {Hayashi}\ \emph
  {et~al.}(2007{\natexlab{b}})\citenamefont {Hayashi}, \citenamefont {Thomas},
  \citenamefont {Rettner}, \citenamefont {Moriya},\ and\ \citenamefont
  {Parkin}}]{bib-HAY2007}%
  \BibitemOpen
  \bibfield  {author} {\bibinfo {author} {\bibfnamefont {M.}~\bibnamefont
  {Hayashi}}, \bibinfo {author} {\bibfnamefont {L.}~\bibnamefont {Thomas}},
  \bibinfo {author} {\bibfnamefont {C.}~\bibnamefont {Rettner}}, \bibinfo
  {author} {\bibfnamefont {R.}~\bibnamefont {Moriya}},\ and\ \bibinfo {author}
  {\bibfnamefont {S.~S.~P.}\ \bibnamefont {Parkin}},\ }\emph {\bibinfo {title}
  {Direct observation of the coherent precession of magnetic domain walls
  propagating along permalloy nanowires}},\ \href
  {https://doi.org/10.1038/nphys464} {\bibfield  {journal} {\bibinfo  {journal}
  {\NatPhys}\ }\textbf {\bibinfo {volume} {3}},\ \bibinfo {pages} {21}
  (\bibinfo {year} {2007}{\natexlab{b}})}\BibitemShut {NoStop}%
\bibitem [{\citenamefont {Thiaville}\ \emph {et~al.}(2012)\citenamefont
  {Thiaville}, \citenamefont {Rohart}, \citenamefont {Jué}, \citenamefont
  {Cros},\ and\ \citenamefont {Fert}}]{bib-THI2012}%
  \BibitemOpen
  \bibfield  {author} {\bibinfo {author} {\bibfnamefont {A.}~\bibnamefont
  {Thiaville}}, \bibinfo {author} {\bibfnamefont {S.}~\bibnamefont {Rohart}},
  \bibinfo {author} {\bibfnamefont {E.}~\bibnamefont {Jué}}, \bibinfo {author}
  {\bibfnamefont {V.}~\bibnamefont {Cros}},\ and\ \bibinfo {author}
  {\bibfnamefont {A.}~\bibnamefont {Fert}},\ }\emph {\bibinfo {title} {Dynamics
  of {Dzyaloshinskii} domain walls in ultrathin magnetic films}},\ \href
  {https://doi.org/10.1209/0295-5075/100/57002} {\bibfield  {journal} {\bibinfo
   {journal} {\epl}\ }\textbf {\bibinfo {volume} {100}},\ \bibinfo {pages}
  {57002} (\bibinfo {year} {2012})}\BibitemShut {NoStop}%
\bibitem [{\citenamefont {Kim}\ \emph {et~al.}(2017)\citenamefont {Kim},
  \citenamefont {Kim}, \citenamefont {Hirata}, \citenamefont {Oh},
  \citenamefont {Tono}, \citenamefont {Kim}, \citenamefont {Okuno},
  \citenamefont {Ham}, \citenamefont {Kim}, \citenamefont {Go}, \citenamefont
  {Tserkovnyak}, \citenamefont {Tsukamoto}, \citenamefont {Moriyama},
  \citenamefont {Lee},\ and\ \citenamefont {Ono}}]{bib-KIM2017c}%
  \BibitemOpen
  \bibfield  {author} {\bibinfo {author} {\bibfnamefont {K.-J.}\ \bibnamefont
  {Kim}}, \bibinfo {author} {\bibfnamefont {S.~K.}\ \bibnamefont {Kim}},
  \bibinfo {author} {\bibfnamefont {Y.}~\bibnamefont {Hirata}}, \bibinfo
  {author} {\bibfnamefont {S.-H.}\ \bibnamefont {Oh}}, \bibinfo {author}
  {\bibfnamefont {T.}~\bibnamefont {Tono}}, \bibinfo {author} {\bibfnamefont
  {D.-H.}\ \bibnamefont {Kim}}, \bibinfo {author} {\bibfnamefont
  {T.}~\bibnamefont {Okuno}}, \bibinfo {author} {\bibfnamefont {W.~S.}\
  \bibnamefont {Ham}}, \bibinfo {author} {\bibfnamefont {S.}~\bibnamefont
  {Kim}}, \bibinfo {author} {\bibfnamefont {G.}~\bibnamefont {Go}}, \bibinfo
  {author} {\bibfnamefont {Y.}~\bibnamefont {Tserkovnyak}}, \bibinfo {author}
  {\bibfnamefont {A.}~\bibnamefont {Tsukamoto}}, \bibinfo {author}
  {\bibfnamefont {T.}~\bibnamefont {Moriyama}}, \bibinfo {author}
  {\bibfnamefont {K.-J.}\ \bibnamefont {Lee}},\ and\ \bibinfo {author}
  {\bibfnamefont {T.}~\bibnamefont {Ono}},\ }\emph {\bibinfo {title} {Fast
  domain wall motion in the vicinity of the angular momentum compensation
  temperature of~ferrimagnets}},\ \href {https://doi.org/10.1038/nmat4990}
  {\bibfield  {journal} {\bibinfo  {journal} {\NatMater}\ }\textbf {\bibinfo
  {volume} {16}},\ \bibinfo {pages} {1187} (\bibinfo {year}
  {2017})}\BibitemShut {NoStop}%
\bibitem [{\citenamefont {Caretta}\ \emph {et~al.}(2018)\citenamefont
  {Caretta}, \citenamefont {Mann}, \citenamefont {Büttner}, \citenamefont
  {Ueda}, \citenamefont {Pfau}, \citenamefont {Günther}, \citenamefont
  {Hessing}, \citenamefont {Churikova}, \citenamefont {Klose}, \citenamefont
  {Schneider}, \citenamefont {Engel}, \citenamefont {Marcus}, \citenamefont
  {Bono}, \citenamefont {Bagschik}, \citenamefont {Eisebitt},\ and\
  \citenamefont {Beach}}]{bib-CAR2018}%
  \BibitemOpen
  \bibfield  {author} {\bibinfo {author} {\bibfnamefont {L.}~\bibnamefont
  {Caretta}}, \bibinfo {author} {\bibfnamefont {M.}~\bibnamefont {Mann}},
  \bibinfo {author} {\bibfnamefont {F.}~\bibnamefont {Büttner}}, \bibinfo
  {author} {\bibfnamefont {K.}~\bibnamefont {Ueda}}, \bibinfo {author}
  {\bibfnamefont {B.}~\bibnamefont {Pfau}}, \bibinfo {author} {\bibfnamefont
  {C.~M.}\ \bibnamefont {Günther}}, \bibinfo {author} {\bibfnamefont
  {P.}~\bibnamefont {Hessing}}, \bibinfo {author} {\bibfnamefont
  {A.}~\bibnamefont {Churikova}}, \bibinfo {author} {\bibfnamefont
  {C.}~\bibnamefont {Klose}}, \bibinfo {author} {\bibfnamefont
  {M.}~\bibnamefont {Schneider}}, \bibinfo {author} {\bibfnamefont
  {D.}~\bibnamefont {Engel}}, \bibinfo {author} {\bibfnamefont
  {C.}~\bibnamefont {Marcus}}, \bibinfo {author} {\bibfnamefont
  {D.}~\bibnamefont {Bono}}, \bibinfo {author} {\bibfnamefont {K.}~\bibnamefont
  {Bagschik}}, \bibinfo {author} {\bibfnamefont {S.}~\bibnamefont {Eisebitt}},\
  and\ \bibinfo {author} {\bibfnamefont {G.~S.~D.}\ \bibnamefont {Beach}},\
  }\emph {\bibinfo {title} {Fast current-driven domain walls and small
  skyrmions in a compensated ferrimagnet}},\ \href
  {https://doi.org/10.1038/s41565-018-0255-3} {\bibfield  {journal} {\bibinfo
  {journal} {\NatNanotech}\ }\textbf {\bibinfo {volume} {13}},\ \bibinfo
  {pages} {1154} (\bibinfo {year} {2018})}\BibitemShut {NoStop}%
\bibitem [{\citenamefont {Fernandez-Pacheco}\ \emph {et~al.}(2017)\citenamefont
  {Fernandez-Pacheco}, \citenamefont {Streubel}, \citenamefont {Fruchart},
  \citenamefont {Hertel}, \citenamefont {Fischer},\ and\ \citenamefont
  {Cowburn}}]{bib-FRU2017b}%
  \BibitemOpen
  \bibfield  {author} {\bibinfo {author} {\bibfnamefont {A.}~\bibnamefont
  {Fernandez-Pacheco}}, \bibinfo {author} {\bibfnamefont {R.}~\bibnamefont
  {Streubel}}, \bibinfo {author} {\bibfnamefont {O.}~\bibnamefont {Fruchart}},
  \bibinfo {author} {\bibfnamefont {R.}~\bibnamefont {Hertel}}, \bibinfo
  {author} {\bibfnamefont {P.}~\bibnamefont {Fischer}},\ and\ \bibinfo {author}
  {\bibfnamefont {R.~P.}\ \bibnamefont {Cowburn}},\ }\emph {\bibinfo {title}
  {Three-dimensional magnetism}},\ \href {https://doi.org/10.1038/ncomms15756}
  {\bibfield  {journal} {\bibinfo  {journal} {\NatComm}\ }\textbf {\bibinfo
  {volume} {8}},\ \bibinfo {pages} {15756} (\bibinfo {year}
  {2017})}\BibitemShut {NoStop}%
\bibitem [{\citenamefont {Streubel}\ \emph {et~al.}(2021)\citenamefont
  {Streubel}, \citenamefont {Tsymbal},\ and\ \citenamefont
  {Fischer}}]{bib-STR2021}%
  \BibitemOpen
  \bibfield  {author} {\bibinfo {author} {\bibfnamefont {R.}~\bibnamefont
  {Streubel}}, \bibinfo {author} {\bibfnamefont {E.~Y.}\ \bibnamefont
  {Tsymbal}},\ and\ \bibinfo {author} {\bibfnamefont {P.}~\bibnamefont
  {Fischer}},\ }\emph {\bibinfo {title} {Magnetism in curved geometries}},\
  \href {https://doi.org/10.1063/5.0054025} {\bibfield  {journal} {\bibinfo
  {journal} {\jap}\ }\textbf {\bibinfo {volume} {129}},\ \bibinfo {pages}
  {210902} (\bibinfo {year} {2021})}\BibitemShut {NoStop}%
\bibitem [{\citenamefont {Makarov}\ and\ \citenamefont
  {Sheka}(2022)}]{bib-MAK2022}%
  \BibitemOpen
  \bibinfo {editor} {\bibfnamefont {D.}~\bibnamefont {Makarov}}\ and\ \bibinfo
  {editor} {\bibfnamefont {D.~D.}\ \bibnamefont {Sheka}},\ eds.,\ \href
  {https://doi.org/10.1007/978-3-031-09086-8} {\emph {\bibinfo {title}
  {Curvilinear micromagnetism, from fundamentals to applications}}},\ \bibinfo
  {series} {Topics in Applied Physics}, Vol.\ \bibinfo {volume} {146}\
  (\bibinfo  {publisher} {Springer International Publishing},\ \bibinfo {year}
  {2022})\BibitemShut {NoStop}%
\bibitem [{\citenamefont {Pylypovskyi}\ \emph {et~al.}(2015)\citenamefont
  {Pylypovskyi}, \citenamefont {Kravchuk}, \citenamefont {Sheka}, \citenamefont
  {Makarov}, \citenamefont {Schmidt},\ and\ \citenamefont
  {Gaididei}}]{bib-PYL2015}%
  \BibitemOpen
  \bibfield  {author} {\bibinfo {author} {\bibfnamefont {O.~V.}\ \bibnamefont
  {Pylypovskyi}}, \bibinfo {author} {\bibfnamefont {V.~P.}\ \bibnamefont
  {Kravchuk}}, \bibinfo {author} {\bibfnamefont {D.~D.}\ \bibnamefont {Sheka}},
  \bibinfo {author} {\bibfnamefont {D.}~\bibnamefont {Makarov}}, \bibinfo
  {author} {\bibfnamefont {O.~G.}\ \bibnamefont {Schmidt}},\ and\ \bibinfo
  {author} {\bibfnamefont {Y.}~\bibnamefont {Gaididei}},\ }\emph {\bibinfo
  {title} {Coupling of chiralities in spin and physical spaces: The möbius
  ring as a case study}},\ \href
  {https://doi.org/10.1103/physrevlett.114.197204} {\bibfield  {journal}
  {\bibinfo  {journal} {\prl}\ }\textbf {\bibinfo {volume} {114}},\ \bibinfo
  {pages} {197204} (\bibinfo {year} {2015})}\BibitemShut {NoStop}%
\bibitem [{\citenamefont {Hertel}(2016)}]{bib-HER2016}%
  \BibitemOpen
  \bibfield  {author} {\bibinfo {author} {\bibfnamefont {R.}~\bibnamefont
  {Hertel}},\ }\emph {\bibinfo {title} {Ultrafast domain wall dynamics in
  magnetic nanotubes and nanowires}},\ \href
  {https://doi.org/10.1088/0953-8984/28/48/483002} {\bibfield  {journal}
  {\bibinfo  {journal} {\jpcm}\ }\textbf {\bibinfo {volume} {28}},\ \bibinfo
  {pages} {483002} (\bibinfo {year} {2016})}\BibitemShut {NoStop}%
\bibitem [{\citenamefont {Yan}\ \emph {et~al.}(2011)\citenamefont {Yan},
  \citenamefont {Andreas}, \citenamefont {Kakay}, \citenamefont
  {Garcia-Sanchez},\ and\ \citenamefont {Hertel}}]{bib-YAN2011b}%
  \BibitemOpen
  \bibfield  {author} {\bibinfo {author} {\bibfnamefont {M.}~\bibnamefont
  {Yan}}, \bibinfo {author} {\bibfnamefont {C.}~\bibnamefont {Andreas}},
  \bibinfo {author} {\bibfnamefont {A.}~\bibnamefont {Kakay}}, \bibinfo
  {author} {\bibfnamefont {F.}~\bibnamefont {Garcia-Sanchez}},\ and\ \bibinfo
  {author} {\bibfnamefont {R.}~\bibnamefont {Hertel}},\ }\emph {\bibinfo
  {title} {Fast domain wall dynamics in magnetic nanotubes: Suppression of
  {Walker} breakdown and {Cherenkov}-like spin wave emission}},\ \href
  {https://doi.org/{10.1063/1.3643037}} {\bibfield  {journal} {\bibinfo
  {journal} {\apl}\ }\textbf {\bibinfo {volume} {99}},\ \bibinfo {pages}
  {122505} (\bibinfo {year} {2011})}\BibitemShut {NoStop}%
\bibitem [{\citenamefont {Schöbitz}\ \emph {et~al.}(2019)\citenamefont
  {Schöbitz}, \citenamefont {{De Riz}}, \citenamefont {Martin}, \citenamefont
  {Bochmann}, \citenamefont {Thirion}, \citenamefont {Vogel}, \citenamefont
  {Foerster}, \citenamefont {Aballe}, \citenamefont {Mente\c{s}}, \citenamefont
  {Locatelli}, \citenamefont {Genuzio}, \citenamefont {{Le Denmat}},
  \citenamefont {Cagnon}, \citenamefont {Toussaint}, \citenamefont {Gusakova},
  \citenamefont {Bachmann},\ and\ \citenamefont {Fruchart}}]{bib-FRU2019b}%
  \BibitemOpen
  \bibfield  {author} {\bibinfo {author} {\bibfnamefont {M.}~\bibnamefont
  {Schöbitz}}, \bibinfo {author} {\bibfnamefont {A.}~\bibnamefont {{De Riz}}},
  \bibinfo {author} {\bibfnamefont {S.}~\bibnamefont {Martin}}, \bibinfo
  {author} {\bibfnamefont {S.}~\bibnamefont {Bochmann}}, \bibinfo {author}
  {\bibfnamefont {C.}~\bibnamefont {Thirion}}, \bibinfo {author} {\bibfnamefont
  {J.}~\bibnamefont {Vogel}}, \bibinfo {author} {\bibfnamefont
  {M.}~\bibnamefont {Foerster}}, \bibinfo {author} {\bibfnamefont
  {L.}~\bibnamefont {Aballe}}, \bibinfo {author} {\bibfnamefont {T.~O.}\
  \bibnamefont {Mente\c{s}}}, \bibinfo {author} {\bibfnamefont
  {A.}~\bibnamefont {Locatelli}}, \bibinfo {author} {\bibfnamefont
  {F.}~\bibnamefont {Genuzio}}, \bibinfo {author} {\bibfnamefont
  {S.}~\bibnamefont {{Le Denmat}}}, \bibinfo {author} {\bibfnamefont
  {L.}~\bibnamefont {Cagnon}}, \bibinfo {author} {\bibfnamefont
  {J.}~\bibnamefont {Toussaint}}, \bibinfo {author} {\bibfnamefont
  {D.}~\bibnamefont {Gusakova}}, \bibinfo {author} {\bibfnamefont
  {J.}~\bibnamefont {Bachmann}},\ and\ \bibinfo {author} {\bibfnamefont
  {O.}~\bibnamefont {Fruchart}},\ }\emph {\bibinfo {title} {Fast domain walls
  governed by topology and \oe sted fields in cylindrical magnetic
  nanowires}},\ \href {https://doi.org/10.1103/PhysRevLett.123.217201}
  {\bibfield  {journal} {\bibinfo  {journal} {\prl}\ }\textbf {\bibinfo
  {volume} {123}},\ \bibinfo {pages} {217201} (\bibinfo {year}
  {2019})}\BibitemShut {NoStop}%
\bibitem [{\citenamefont {Bran}\ \emph {et~al.}(2023)\citenamefont {Bran},
  \citenamefont {Fernandez-Roldan}, \citenamefont {Moreno}, \citenamefont
  {Rodríguez}, \citenamefont {del Real}, \citenamefont {Asenjo}, \citenamefont
  {Saugar}, \citenamefont {Marqués-Marchán}, \citenamefont {Mohammed},
  \citenamefont {Foerster}, \citenamefont {Aballe}, \citenamefont {Kosel},
  \citenamefont {Vazquez},\ and\ \citenamefont
  {Chubykalo-Fesenko}}]{bib-BRA2023}%
  \BibitemOpen
  \bibfield  {author} {\bibinfo {author} {\bibfnamefont {C.}~\bibnamefont
  {Bran}}, \bibinfo {author} {\bibfnamefont {J.~A.}\ \bibnamefont
  {Fernandez-Roldan}}, \bibinfo {author} {\bibfnamefont {J.~A.}\ \bibnamefont
  {Moreno}}, \bibinfo {author} {\bibfnamefont {A.~F.}\ \bibnamefont
  {Rodríguez}}, \bibinfo {author} {\bibfnamefont {R.~P.}\ \bibnamefont {del
  Real}}, \bibinfo {author} {\bibfnamefont {A.}~\bibnamefont {Asenjo}},
  \bibinfo {author} {\bibfnamefont {E.}~\bibnamefont {Saugar}}, \bibinfo
  {author} {\bibfnamefont {J.}~\bibnamefont {Marqués-Marchán}}, \bibinfo
  {author} {\bibfnamefont {H.}~\bibnamefont {Mohammed}}, \bibinfo {author}
  {\bibfnamefont {M.}~\bibnamefont {Foerster}}, \bibinfo {author}
  {\bibfnamefont {L.}~\bibnamefont {Aballe}}, \bibinfo {author} {\bibfnamefont
  {J.}~\bibnamefont {Kosel}}, \bibinfo {author} {\bibfnamefont
  {M.}~\bibnamefont {Vazquez}},\ and\ \bibinfo {author} {\bibfnamefont
  {O.}~\bibnamefont {Chubykalo-Fesenko}},\ }\emph {\bibinfo {title} {Domain
  wall propagation and pinning induced by current pulses in cylindrical
  modulated nanowires}},\ \href {https://doi.org/10.1039/d3nr00455d} {\bibfield
   {journal} {\bibinfo  {journal} {\Nanoscale}\ }\textbf {\bibinfo {volume}
  {18}},\ \bibinfo {pages} {8387} (\bibinfo {year} {2023})},\ \Eprint
  {https://arxiv.org/abs/2210.01480} {arXiv:2210.01480 [cond-mat.mtrl-sci]}
  \BibitemShut {NoStop}%
\bibitem [{\citenamefont {Ot{\'{a}}lora}\ \emph {et~al.}(2016)\citenamefont
  {Ot{\'{a}}lora}, \citenamefont {Yan}, \citenamefont {Schultheiss},
  \citenamefont {Hertel},\ and\ \citenamefont {K{\'{a}}kay}}]{bib-OTA2016}%
  \BibitemOpen
  \bibfield  {author} {\bibinfo {author} {\bibfnamefont {J.~A.}\ \bibnamefont
  {Ot{\'{a}}lora}}, \bibinfo {author} {\bibfnamefont {M.}~\bibnamefont {Yan}},
  \bibinfo {author} {\bibfnamefont {H.}~\bibnamefont {Schultheiss}}, \bibinfo
  {author} {\bibfnamefont {R.}~\bibnamefont {Hertel}},\ and\ \bibinfo {author}
  {\bibfnamefont {A.}~\bibnamefont {K{\'{a}}kay}},\ }\emph {\bibinfo {title}
  {Curvature-induced asymmetric spin-wave dispersion}},\ \href
  {https://doi.org/10.1103/physrevlett.117.227203} {\bibfield  {journal}
  {\bibinfo  {journal} {\prl}\ }\textbf {\bibinfo {volume} {117}},\ \bibinfo
  {pages} {227203} (\bibinfo {year} {2016})}\BibitemShut {NoStop}%
\bibitem [{\citenamefont {Körber}\ \emph {et~al.}(2022)\citenamefont
  {Körber}, \citenamefont {K{\'{e}}zsm{\'{a}}rki},\ and\ \citenamefont
  {K{\'{a}}kay}}]{bib-KOE2022}%
  \BibitemOpen
  \bibfield  {author} {\bibinfo {author} {\bibfnamefont {L.}~\bibnamefont
  {Körber}}, \bibinfo {author} {\bibfnamefont {I.}~\bibnamefont
  {K{\'{e}}zsm{\'{a}}rki}},\ and\ \bibinfo {author} {\bibfnamefont
  {A.}~\bibnamefont {K{\'{a}}kay}},\ }\emph {\bibinfo {title} {Mode splitting
  of spin waves in magnetic nanotubes with discrete symmetries}},\ \href
  {https://doi.org/10.1103/physrevb.105.184435} {\bibfield  {journal} {\bibinfo
   {journal} {\prb}\ }\textbf {\bibinfo {volume} {105}},\ \bibinfo {pages}
  {184435} (\bibinfo {year} {2022})}\BibitemShut {NoStop}%
\bibitem [{\citenamefont {Stancil}\ and\ \citenamefont
  {Prabhakar}(2009)}]{bib-STA2009}%
  \BibitemOpen
  \bibfield  {author} {\bibinfo {author} {\bibfnamefont {D.~D.}\ \bibnamefont
  {Stancil}}\ and\ \bibinfo {author} {\bibfnamefont {A.}~\bibnamefont
  {Prabhakar}},\ }\href {https://doi.org/10.1007/978-0-387-77865-5} {\emph
  {\bibinfo {title} {Spin waves. Theory and applications}}}\ (\bibinfo
  {publisher} {Springer},\ \bibinfo {address} {New-York},\ \bibinfo {year}
  {2009})\BibitemShut {NoStop}%
\bibitem [{\citenamefont {Ruiz-G{\'{o}}mez}\ \emph {et~al.}(2018)\citenamefont
  {Ruiz-G{\'{o}}mez}, \citenamefont {Foerster}, \citenamefont {Aballe},
  \citenamefont {Proenca}, \citenamefont {Lucas}, \citenamefont {Prieto},
  \citenamefont {Mascaraque}, \citenamefont {de~la Figuera}, \citenamefont
  {Quesada},\ and\ \citenamefont {Pérez}}]{bib-RUI2018}%
  \BibitemOpen
  \bibfield  {author} {\bibinfo {author} {\bibfnamefont {S.}~\bibnamefont
  {Ruiz-G{\'{o}}mez}}, \bibinfo {author} {\bibfnamefont {M.}~\bibnamefont
  {Foerster}}, \bibinfo {author} {\bibfnamefont {L.}~\bibnamefont {Aballe}},
  \bibinfo {author} {\bibfnamefont {M.~P.}\ \bibnamefont {Proenca}}, \bibinfo
  {author} {\bibfnamefont {I.}~\bibnamefont {Lucas}}, \bibinfo {author}
  {\bibfnamefont {J.~L.}\ \bibnamefont {Prieto}}, \bibinfo {author}
  {\bibfnamefont {A.}~\bibnamefont {Mascaraque}}, \bibinfo {author}
  {\bibfnamefont {J.}~\bibnamefont {de~la Figuera}}, \bibinfo {author}
  {\bibfnamefont {A.}~\bibnamefont {Quesada}},\ and\ \bibinfo {author}
  {\bibfnamefont {L.}~\bibnamefont {Pérez}},\ }\emph {\bibinfo {title}
  {Observation of a topologically protected state in a magnetic domain wall
  stabilized by a ferromagnetic chemical barrier}},\ \href
  {https://doi.org/10.1038/s41598-018-35039-6} {\bibfield  {journal} {\bibinfo
  {journal} {\SciRep}\ }\textbf {\bibinfo {volume} {8}},\ \bibinfo {pages}
  {16695} (\bibinfo {year} {2018})}\BibitemShut {NoStop}%
\bibitem [{\citenamefont {Fernandez-Roldan}\ \emph {et~al.}(2022)\citenamefont
  {Fernandez-Roldan}, \citenamefont {Bran}, \citenamefont {Asenjo},
  \citenamefont {Vázquez}, \citenamefont {Sorrentino}, \citenamefont {Ferrer},
  \citenamefont {Chubykalo-Fesenko},\ and\ \citenamefont {del
  Real}}]{bib-FER2022}%
  \BibitemOpen
  \bibfield  {author} {\bibinfo {author} {\bibfnamefont {J.~A.}\ \bibnamefont
  {Fernandez-Roldan}}, \bibinfo {author} {\bibfnamefont {C.}~\bibnamefont
  {Bran}}, \bibinfo {author} {\bibfnamefont {A.}~\bibnamefont {Asenjo}},
  \bibinfo {author} {\bibfnamefont {M.}~\bibnamefont {Vázquez}}, \bibinfo
  {author} {\bibfnamefont {A.}~\bibnamefont {Sorrentino}}, \bibinfo {author}
  {\bibfnamefont {S.}~\bibnamefont {Ferrer}}, \bibinfo {author} {\bibfnamefont
  {O.}~\bibnamefont {Chubykalo-Fesenko}},\ and\ \bibinfo {author}
  {\bibfnamefont {R.~P.}\ \bibnamefont {del Real}},\ }\emph {\bibinfo {title}
  {Spatial magnetic imaging of non-axially symmetric vortex domains in
  cylindrical nanowire by transmission x-ray microscopy}},\ \href
  {https://doi.org/10.1039/d2nr03228g} {\bibfield  {journal} {\bibinfo
  {journal} {\Nanoscale}\ }\textbf {\bibinfo {volume} {14}},\ \bibinfo {pages}
  {13661} (\bibinfo {year} {2022})}\BibitemShut {NoStop}%
\bibitem [{\citenamefont {Coehoorn}\ \emph {et~al.}(1988)\citenamefont
  {Coehoorn}, \citenamefont {{de Mooij}}, \citenamefont {Duchateau},\ and\
  \citenamefont {Buschow}}]{bib-COE1988}%
  \BibitemOpen
  \bibfield  {author} {\bibinfo {author} {\bibfnamefont {R.}~\bibnamefont
  {Coehoorn}}, \bibinfo {author} {\bibfnamefont {D.~B.}\ \bibnamefont {{de
  Mooij}}}, \bibinfo {author} {\bibfnamefont {J.~P. W.~B.}\ \bibnamefont
  {Duchateau}},\ and\ \bibinfo {author} {\bibfnamefont {L.~H.~J.}\ \bibnamefont
  {Buschow}},\ }\emph {\bibinfo {title} {Novel permanent magnetic materials
  made by rapid quenching}},\ \bibfield  {journal} {\bibinfo  {journal}
  {\JourPhys}\ }\textbf {\bibinfo {volume} {49 C-8}},\ \href
  {https://doi.org/10.1051/jphyscol:19888304} {10.1051/jphyscol:19888304}
  (\bibinfo {year} {1988})\BibitemShut {NoStop}%
\bibitem [{\citenamefont {Kneller}(1991)}]{bib-KNE1991}%
  \BibitemOpen
  \bibfield  {author} {\bibinfo {author} {\bibfnamefont {E.~F.}\ \bibnamefont
  {Kneller}},\ }\emph {\bibinfo {title} {The exchange-spring magnet: a new
  material principle for permanent magnets}},\ \href
  {https://doi.org/10.1109/20.102931} {\bibfield  {journal} {\bibinfo
  {journal} {\IEEEtm}\ }\textbf {\bibinfo {volume} {27}},\ \bibinfo {pages}
  {3588} (\bibinfo {year} {1991})}\BibitemShut {NoStop}%
\bibitem [{\citenamefont {Skomski}\ and\ \citenamefont
  {Coey}(1993)}]{bib-SKO1993}%
  \BibitemOpen
  \bibfield  {author} {\bibinfo {author} {\bibfnamefont {R.}~\bibnamefont
  {Skomski}}\ and\ \bibinfo {author} {\bibfnamefont {J.~M.~D.}\ \bibnamefont
  {Coey}},\ }\emph {\bibinfo {title} {Giant energy product in nanostructured
  two-phase magnets}},\ \href {https://doi.org/10.1103/PhysRevB.48.15812}
  {\bibfield  {journal} {\bibinfo  {journal} {\prb}\ }\textbf {\bibinfo
  {volume} {48}},\ \bibinfo {pages} {15812} (\bibinfo {year}
  {1993})}\BibitemShut {NoStop}%
\bibitem [{\citenamefont {Stöhr}\ \emph {et~al.}(1998)\citenamefont {Stöhr},
  \citenamefont {Padmore}, \citenamefont {Anders}, \citenamefont {Stammler},\
  and\ \citenamefont {Scheinfein}}]{bib-STO1998}%
  \BibitemOpen
  \bibfield  {author} {\bibinfo {author} {\bibfnamefont {J.}~\bibnamefont
  {Stöhr}}, \bibinfo {author} {\bibfnamefont {H.~A.}\ \bibnamefont {Padmore}},
  \bibinfo {author} {\bibfnamefont {S.}~\bibnamefont {Anders}}, \bibinfo
  {author} {\bibfnamefont {T.}~\bibnamefont {Stammler}},\ and\ \bibinfo
  {author} {\bibfnamefont {M.~R.}\ \bibnamefont {Scheinfein}},\ }\emph
  {\bibinfo {title} {Principles of x-ray magnetic dichroism
  spectromicroscopy}},\ \href {https://doi.org/10.1142/s0218625x98001638}
  {\bibfield  {journal} {\bibinfo  {journal} {\srl}\ }\textbf {\bibinfo
  {volume} {05}},\ \bibinfo {pages} {1297} (\bibinfo {year}
  {1998})}\BibitemShut {NoStop}%
\bibitem [{\citenamefont {Bushida}\ \emph {et~al.}(1994)\citenamefont
  {Bushida}, \citenamefont {Noda}, \citenamefont {Panina}, \citenamefont
  {Yoshida}, \citenamefont {Uchiyama},\ and\ \citenamefont
  {Mohri}}]{bib-BUS1994}%
  \BibitemOpen
  \bibfield  {author} {\bibinfo {author} {\bibfnamefont {K.}~\bibnamefont
  {Bushida}}, \bibinfo {author} {\bibfnamefont {M.}~\bibnamefont {Noda}},
  \bibinfo {author} {\bibfnamefont {L.}~\bibnamefont {Panina}}, \bibinfo
  {author} {\bibfnamefont {H.}~\bibnamefont {Yoshida}}, \bibinfo {author}
  {\bibfnamefont {T.}~\bibnamefont {Uchiyama}},\ and\ \bibinfo {author}
  {\bibfnamefont {K.}~\bibnamefont {Mohri}},\ }\emph {\bibinfo {title}
  {Magneto-impedance element using amorphous micro wire}},\ \href
  {https://doi.org/10.1109/tjmj.1994.4565950} {\bibfield  {journal} {\bibinfo
  {journal} {\IEEEtmj}\ }\textbf {\bibinfo {volume} {9}},\ \bibinfo {pages} {7}
  (\bibinfo {year} {1994})}\BibitemShut {NoStop}%
\bibitem [{\citenamefont {Panina}\ and\ \citenamefont
  {Mohri}(1994)}]{bib-PAN1994}%
  \BibitemOpen
  \bibfield  {author} {\bibinfo {author} {\bibfnamefont {L.~V.}\ \bibnamefont
  {Panina}}\ and\ \bibinfo {author} {\bibfnamefont {K.}~\bibnamefont {Mohri}},\
  }\emph {\bibinfo {title} {Magneto-impedance effect in amorphous wires}},\
  \href {https://doi.org/10.1063/1.112104} {\bibfield  {journal} {\bibinfo
  {journal} {Applied Physics Letters}\ }\textbf {\bibinfo {volume} {65}},\
  \bibinfo {pages} {1189} (\bibinfo {year} {1994})}\BibitemShut {NoStop}%
\bibitem [{\citenamefont {Álvaro Gómez}\ \emph {et~al.}(2022)\citenamefont
  {Álvaro Gómez}, \citenamefont {Ruiz-Gómez}, \citenamefont
  {Fernández-González}, \citenamefont {Schöbitz}, \citenamefont {Mille},
  \citenamefont {Hurst}, \citenamefont {Tiwari}, \citenamefont {{de Riz}},
  \citenamefont {Andersen}, \citenamefont {Bachmann}, \citenamefont {Cagnon},
  \citenamefont {Foerster}, \citenamefont {Aballe}, \citenamefont {Belkhou},
  \citenamefont {Toussaint}, \citenamefont {Thirion}, \citenamefont
  {Masseboeuf}, \citenamefont {Gusakova}, \citenamefont {Pérez},\ and\
  \citenamefont {Fruchart}}]{bib-FRU2022}%
  \BibitemOpen
  \bibfield  {author} {\bibinfo {author} {\bibfnamefont {L.}~\bibnamefont
  {Álvaro Gómez}}, \bibinfo {author} {\bibfnamefont {S.}~\bibnamefont
  {Ruiz-Gómez}}, \bibinfo {author} {\bibfnamefont {C.}~\bibnamefont
  {Fernández-González}}, \bibinfo {author} {\bibfnamefont {M.}~\bibnamefont
  {Schöbitz}}, \bibinfo {author} {\bibfnamefont {N.}~\bibnamefont {Mille}},
  \bibinfo {author} {\bibfnamefont {J.}~\bibnamefont {Hurst}}, \bibinfo
  {author} {\bibfnamefont {D.}~\bibnamefont {Tiwari}}, \bibinfo {author}
  {\bibfnamefont {A.}~\bibnamefont {{de Riz}}}, \bibinfo {author}
  {\bibfnamefont {I.~M.}\ \bibnamefont {Andersen}}, \bibinfo {author}
  {\bibfnamefont {J.}~\bibnamefont {Bachmann}}, \bibinfo {author}
  {\bibfnamefont {L.}~\bibnamefont {Cagnon}}, \bibinfo {author} {\bibfnamefont
  {M.}~\bibnamefont {Foerster}}, \bibinfo {author} {\bibfnamefont
  {L.}~\bibnamefont {Aballe}}, \bibinfo {author} {\bibfnamefont
  {R.}~\bibnamefont {Belkhou}}, \bibinfo {author} {\bibfnamefont {J.~C.}\
  \bibnamefont {Toussaint}}, \bibinfo {author} {\bibfnamefont {C.}~\bibnamefont
  {Thirion}}, \bibinfo {author} {\bibfnamefont {A.}~\bibnamefont {Masseboeuf}},
  \bibinfo {author} {\bibfnamefont {D.}~\bibnamefont {Gusakova}}, \bibinfo
  {author} {\bibfnamefont {L.}~\bibnamefont {Pérez}},\ and\ \bibinfo {author}
  {\bibfnamefont {O.}~\bibnamefont {Fruchart}},\ }\emph {\bibinfo {title}
  {Micromagnetics of magnetic chemical modulations in soft-magnetic cylindrical
  nanowires}},\ \href {https://doi.org/10.1103/PhysRevB.106.054433} {\bibfield
  {journal} {\bibinfo  {journal} {\prb}\ }\textbf {\bibinfo {volume} {106}},\
  \bibinfo {pages} {054433} (\bibinfo {year} {2022})}\BibitemShut {NoStop}%
\bibitem [{\citenamefont {Chizhik}\ \emph {et~al.}(2003)\citenamefont
  {Chizhik}, \citenamefont {Gonzalez}, \citenamefont {Zhukov},\ and\
  \citenamefont {Blanco}}]{bib-CHI2003}%
  \BibitemOpen
  \bibfield  {author} {\bibinfo {author} {\bibfnamefont {A.}~\bibnamefont
  {Chizhik}}, \bibinfo {author} {\bibfnamefont {J.}~\bibnamefont {Gonzalez}},
  \bibinfo {author} {\bibfnamefont {A.}~\bibnamefont {Zhukov}},\ and\ \bibinfo
  {author} {\bibfnamefont {J.~M.}\ \bibnamefont {Blanco}},\ }\emph {\bibinfo
  {title} {Circular magnetic bistability in co-rich amorphous microwires}},\
  \href {https://doi.org/10.1088/0022-3727/36/5/301} {\bibfield  {journal}
  {\bibinfo  {journal} {\jpdap}\ }\textbf {\bibinfo {volume} {36}},\ \bibinfo
  {pages} {419} (\bibinfo {year} {2003})}\BibitemShut {NoStop}%
\bibitem [{\citenamefont {Chizhik}\ \emph {et~al.}(2009)\citenamefont
  {Chizhik}, \citenamefont {Zhukov}, \citenamefont {Stupakiewicz},
  \citenamefont {Maziewski}, \citenamefont {Blanco},\ and\ \citenamefont
  {Gonzalez}}]{bib-CHI2009}%
  \BibitemOpen
  \bibfield  {author} {\bibinfo {author} {\bibfnamefont {A.}~\bibnamefont
  {Chizhik}}, \bibinfo {author} {\bibfnamefont {A.}~\bibnamefont {Zhukov}},
  \bibinfo {author} {\bibfnamefont {A.}~\bibnamefont {Stupakiewicz}}, \bibinfo
  {author} {\bibfnamefont {A.}~\bibnamefont {Maziewski}}, \bibinfo {author}
  {\bibfnamefont {J.}~\bibnamefont {Blanco}},\ and\ \bibinfo {author}
  {\bibfnamefont {J.}~\bibnamefont {Gonzalez}},\ }\emph {\bibinfo {title} {Kerr
  microscopy study of magnetic domain structure changes in amorphous
  microwires}},\ \href {https://doi.org/10.1109/tmag.2009.2024892} {\bibfield
  {journal} {\bibinfo  {journal} {\IEEEtm}\ }\textbf {\bibinfo {volume} {45}},\
  \bibinfo {pages} {4279} (\bibinfo {year} {2009})}\BibitemShut {NoStop}%
\bibitem [{\citenamefont {Zhukova}\ \emph {et~al.}(2018)\citenamefont
  {Zhukova}, \citenamefont {Blanco}, \citenamefont {Chizhik}, \citenamefont
  {Ipatov},\ and\ \citenamefont {Zhukov}}]{bib-ZHU2018}%
  \BibitemOpen
  \bibfield  {author} {\bibinfo {author} {\bibfnamefont {V.}~\bibnamefont
  {Zhukova}}, \bibinfo {author} {\bibfnamefont {J.~M.}\ \bibnamefont {Blanco}},
  \bibinfo {author} {\bibfnamefont {A.}~\bibnamefont {Chizhik}}, \bibinfo
  {author} {\bibfnamefont {M.}~\bibnamefont {Ipatov}},\ and\ \bibinfo {author}
  {\bibfnamefont {A.}~\bibnamefont {Zhukov}},\ }\emph {\bibinfo {title}
  {Ac-current-induced magnetization switching in amorphous microwires}},\
  \bibfield  {journal} {\bibinfo  {journal} {\FrontPhys}\ }\textbf {\bibinfo
  {volume} {13}},\ \href {https://doi.org/10.1007/s11467-017-0722-6}
  {10.1007/s11467-017-0722-6} (\bibinfo {year} {2018})\BibitemShut {NoStop}%
\bibitem [{\citenamefont {Jamet}\ \emph
  {et~al.}(2015{\natexlab{a}})\citenamefont {Jamet}, \citenamefont
  {Rougemaille}, \citenamefont {Toussaint},\ and\ \citenamefont
  {Fruchart}}]{bib-FRU2015b}%
  \BibitemOpen
  \bibfield  {author} {\bibinfo {author} {\bibfnamefont {S.}~\bibnamefont
  {Jamet}}, \bibinfo {author} {\bibfnamefont {N.}~\bibnamefont {Rougemaille}},
  \bibinfo {author} {\bibfnamefont {J.~C.}\ \bibnamefont {Toussaint}},\ and\
  \bibinfo {author} {\bibfnamefont {O.}~\bibnamefont {Fruchart}},\ }\bibinfo
  {title} {Magnetic nano- and microwires: Design, synthesis, properties and
  applications}\ (\bibinfo  {publisher} {Woodhead},\ \bibinfo {year} {2015})\
  Chap.\ \bibinfo {chapter} {Head-to-head domain walls in one-dimensional
  nanostructures: an extended phase diagram ranging from strips to cylindrical
  wires}, pp.\ \bibinfo {pages} {783--811}\BibitemShut {NoStop}%
\bibitem [{\citenamefont {Kuepferling}\ \emph {et~al.}(2023)\citenamefont
  {Kuepferling}, \citenamefont {Casiraghi}, \citenamefont {Soares},
  \citenamefont {Durin}, \citenamefont {Garcia-Sanchez}, \citenamefont {Chen},
  \citenamefont {Back}, \citenamefont {Marrows}, \citenamefont {Tacchi},\ and\
  \citenamefont {Carlotti}}]{bib-KUE2023}%
  \BibitemOpen
  \bibfield  {author} {\bibinfo {author} {\bibfnamefont {M.}~\bibnamefont
  {Kuepferling}}, \bibinfo {author} {\bibfnamefont {A.}~\bibnamefont
  {Casiraghi}}, \bibinfo {author} {\bibfnamefont {G.}~\bibnamefont {Soares}},
  \bibinfo {author} {\bibfnamefont {G.}~\bibnamefont {Durin}}, \bibinfo
  {author} {\bibfnamefont {F.}~\bibnamefont {Garcia-Sanchez}}, \bibinfo
  {author} {\bibfnamefont {L.}~\bibnamefont {Chen}}, \bibinfo {author}
  {\bibfnamefont {C.}~\bibnamefont {Back}}, \bibinfo {author} {\bibfnamefont
  {C.}~\bibnamefont {Marrows}}, \bibinfo {author} {\bibfnamefont
  {S.}~\bibnamefont {Tacchi}},\ and\ \bibinfo {author} {\bibfnamefont
  {G.}~\bibnamefont {Carlotti}},\ }\emph {\bibinfo {title} {Measuring
  interfacial dzyaloshinskii-moriya interaction in ultrathin magnetic films}},\
  \href {https://doi.org/10.1103/revmodphys.95.015003} {\bibfield  {journal}
  {\bibinfo  {journal} {Reviews of Modern Physics}\ }\textbf {\bibinfo {volume}
  {95}},\ \bibinfo {pages} {015003} (\bibinfo {year} {2023})}\BibitemShut
  {NoStop}%
\bibitem [{bib({\natexlab{a}})}]{bib-FEE}%
  \BibitemOpen
  \href@noop {} {} ({\natexlab{a}}),\ \bibinfo {note}
  {{http://feellgood.neel.cnrs.fr}{http://feellgood.neel.cnrs.fr}}\BibitemShut
  {NoStop}%
\bibitem [{\citenamefont {Vansteenkiste}\ \emph {et~al.}(2014)\citenamefont
  {Vansteenkiste}, \citenamefont {Leliaert}, \citenamefont {Dvornik},
  \citenamefont {Helsen}, \citenamefont {Garcia-Sanchez},\ and\ \citenamefont
  {Waeyenberge}}]{bib-VAN2014}%
  \BibitemOpen
  \bibfield  {author} {\bibinfo {author} {\bibfnamefont {A.}~\bibnamefont
  {Vansteenkiste}}, \bibinfo {author} {\bibfnamefont {J.}~\bibnamefont
  {Leliaert}}, \bibinfo {author} {\bibfnamefont {M.}~\bibnamefont {Dvornik}},
  \bibinfo {author} {\bibfnamefont {M.}~\bibnamefont {Helsen}}, \bibinfo
  {author} {\bibfnamefont {F.}~\bibnamefont {Garcia-Sanchez}},\ and\ \bibinfo
  {author} {\bibfnamefont {B.~V.}\ \bibnamefont {Waeyenberge}},\ }\emph
  {\bibinfo {title} {The design and verification of {MuMax}3}},\ \href
  {https://doi.org/10.1063/1.4899186} {\bibfield  {journal} {\bibinfo
  {journal} {{AIP} Advances}\ }\textbf {\bibinfo {volume} {4}},\ \bibinfo
  {pages} {107133} (\bibinfo {year} {2014})}\BibitemShut {NoStop}%
\bibitem [{\citenamefont {LaBonte}(1969)}]{bib-LAB1969}%
  \BibitemOpen
  \bibfield  {author} {\bibinfo {author} {\bibfnamefont {A.~E.}\ \bibnamefont
  {LaBonte}},\ }\emph {\bibinfo {title} {Two-dimensional bloch-type domain
  walls in ferromagnetic thin films}},\ \href
  {https://doi.org/10.1063/1.1658014} {\bibfield  {journal} {\bibinfo
  {journal} {\jap}\ }\textbf {\bibinfo {volume} {40}},\ \bibinfo {pages} {2450}
  (\bibinfo {year} {1969})}\BibitemShut {NoStop}%
\bibitem [{\citenamefont {Hubert}(1969)}]{bib-HUB1969}%
  \BibitemOpen
  \bibfield  {author} {\bibinfo {author} {\bibfnamefont {A.}~\bibnamefont
  {Hubert}},\ }\emph {\bibinfo {title} {Stray-field-free magnetization
  configurations}},\ \href@noop {} {\bibfield  {journal} {\bibinfo  {journal}
  {\pss}\ }\textbf {\bibinfo {volume} {32}},\ \bibinfo {pages} {519} (\bibinfo
  {year} {1969})}\BibitemShut {NoStop}%
\bibitem [{\citenamefont {Jamet}\ \emph
  {et~al.}(2015{\natexlab{b}})\citenamefont {Jamet}, \citenamefont {Col},
  \citenamefont {Rougemaille}, \citenamefont {Wartelle}, \citenamefont
  {Locatelli}, \citenamefont {Mente\c{s}}, \citenamefont {Burgos},
  \citenamefont {R.Afid}, \citenamefont {Cagnon}, \citenamefont {Bachmann},
  \citenamefont {Bochmann}, \citenamefont {Fruchart}, ,\ and\ \citenamefont
  {Toussaint}}]{bib-FRU2015c}%
  \BibitemOpen
  \bibfield  {author} {\bibinfo {author} {\bibfnamefont {S.}~\bibnamefont
  {Jamet}}, \bibinfo {author} {\bibfnamefont {S.~D.}\ \bibnamefont {Col}},
  \bibinfo {author} {\bibfnamefont {N.}~\bibnamefont {Rougemaille}}, \bibinfo
  {author} {\bibfnamefont {A.}~\bibnamefont {Wartelle}}, \bibinfo {author}
  {\bibfnamefont {A.}~\bibnamefont {Locatelli}}, \bibinfo {author}
  {\bibfnamefont {T.~O.}\ \bibnamefont {Mente\c{s}}}, \bibinfo {author}
  {\bibfnamefont {B.~S.}\ \bibnamefont {Burgos}}, \bibinfo {author}
  {\bibnamefont {R.Afid}}, \bibinfo {author} {\bibfnamefont {L.}~\bibnamefont
  {Cagnon}}, \bibinfo {author} {\bibfnamefont {J.}~\bibnamefont {Bachmann}},
  \bibinfo {author} {\bibfnamefont {S.}~\bibnamefont {Bochmann}}, \bibinfo
  {author} {\bibfnamefont {O.}~\bibnamefont {Fruchart}}, ,\ and\ \bibinfo
  {author} {\bibfnamefont {J.~C.}\ \bibnamefont {Toussaint}},\ }\emph {\bibinfo
  {title} {Quantitative analysis of shadow x-ray magnetic circular dichroism
  photo-emission electron microscopy}},\ \href
  {https://doi.org/10.1103/PhysRevB.92.144428} {\bibfield  {journal} {\bibinfo
  {journal} {\prb}\ }\textbf {\bibinfo {volume} {92}},\ \bibinfo {pages}
  {144428} (\bibinfo {year} {2015}{\natexlab{b}})}\BibitemShut {NoStop}%
\bibitem [{\citenamefont {Hurst}\ \emph {et~al.}(2021)\citenamefont {Hurst},
  \citenamefont {{De Riz}}, \citenamefont {Sta{\v{n}}o}, \citenamefont
  {Toussaint}, \citenamefont {Fruchart},\ and\ \citenamefont
  {Gusakova}}]{bib-FRU2021c}%
  \BibitemOpen
  \bibfield  {author} {\bibinfo {author} {\bibfnamefont {J.}~\bibnamefont
  {Hurst}}, \bibinfo {author} {\bibfnamefont {A.}~\bibnamefont {{De Riz}}},
  \bibinfo {author} {\bibfnamefont {M.}~\bibnamefont {Sta{\v{n}}o}}, \bibinfo
  {author} {\bibfnamefont {J.-C.}\ \bibnamefont {Toussaint}}, \bibinfo {author}
  {\bibfnamefont {O.}~\bibnamefont {Fruchart}},\ and\ \bibinfo {author}
  {\bibfnamefont {D.}~\bibnamefont {Gusakova}},\ }\emph {\bibinfo {title}
  {Theoretical study of current-induced domain wall motion in magnetic
  nanotubes with azimuthal domains}},\ \href
  {https://doi.org/10.1103/physrevb.103.024434} {\bibfield  {journal} {\bibinfo
   {journal} {\prb}\ }\textbf {\bibinfo {volume} {103}},\ \bibinfo {pages}
  {024434} (\bibinfo {year} {2021})}\BibitemShut {NoStop}%
\bibitem [{\citenamefont {Wartelle}\ \emph {et~al.}(2019)\citenamefont
  {Wartelle}, \citenamefont {Trapp}, \citenamefont {Sta{\v{n}}o}, \citenamefont
  {Thirion}, \citenamefont {Bochmann}, \citenamefont {Bachmann}, \citenamefont
  {Foerster}, \citenamefont {Aballe}, \citenamefont {Mente\c{s}}, \citenamefont
  {Locatelli}, \citenamefont {Sala}, \citenamefont {Cagnon}, \citenamefont
  {Toussaint},\ and\ \citenamefont {Fruchart}}]{bib-FRU2019}%
  \BibitemOpen
  \bibfield  {author} {\bibinfo {author} {\bibfnamefont {A.}~\bibnamefont
  {Wartelle}}, \bibinfo {author} {\bibfnamefont {B.}~\bibnamefont {Trapp}},
  \bibinfo {author} {\bibfnamefont {M.}~\bibnamefont {Sta{\v{n}}o}}, \bibinfo
  {author} {\bibfnamefont {C.}~\bibnamefont {Thirion}}, \bibinfo {author}
  {\bibfnamefont {S.}~\bibnamefont {Bochmann}}, \bibinfo {author}
  {\bibfnamefont {J.}~\bibnamefont {Bachmann}}, \bibinfo {author}
  {\bibfnamefont {M.}~\bibnamefont {Foerster}}, \bibinfo {author}
  {\bibfnamefont {L.}~\bibnamefont {Aballe}}, \bibinfo {author} {\bibfnamefont
  {T.~O.}\ \bibnamefont {Mente\c{s}}}, \bibinfo {author} {\bibfnamefont
  {A.}~\bibnamefont {Locatelli}}, \bibinfo {author} {\bibfnamefont
  {A.}~\bibnamefont {Sala}}, \bibinfo {author} {\bibfnamefont {L.}~\bibnamefont
  {Cagnon}}, \bibinfo {author} {\bibfnamefont {J.}~\bibnamefont {Toussaint}},\
  and\ \bibinfo {author} {\bibfnamefont {O.}~\bibnamefont {Fruchart}},\ }\emph
  {\bibinfo {title} {Bloch-point-mediated topological transformations of
  magnetic domain walls in cylindrical nanowires}},\ \href
  {https://doi.org/10.1103/PhysRevB.99.024433} {\bibfield  {journal} {\bibinfo
  {journal} {\prb}\ }\textbf {\bibinfo {volume} {99}},\ \bibinfo {pages}
  {024433} (\bibinfo {year} {2019})},\ \bibinfo {note}
  {arXiv:1806.10918}\BibitemShut {NoStop}%
\bibitem [{\citenamefont {{De Riz}}\ \emph {et~al.}(2021)\citenamefont {{De
  Riz}}, \citenamefont {Hurst}, \citenamefont {Schöbitz}, \citenamefont
  {Thirion}, \citenamefont {Bachmann}, \citenamefont {Toussaint}, \citenamefont
  {Fruchart},\ and\ \citenamefont {Gusakova}}]{bib-FRU2021}%
  \BibitemOpen
  \bibfield  {author} {\bibinfo {author} {\bibfnamefont {A.}~\bibnamefont {{De
  Riz}}}, \bibinfo {author} {\bibfnamefont {J.}~\bibnamefont {Hurst}}, \bibinfo
  {author} {\bibfnamefont {M.}~\bibnamefont {Schöbitz}}, \bibinfo {author}
  {\bibfnamefont {C.}~\bibnamefont {Thirion}}, \bibinfo {author} {\bibfnamefont
  {J.}~\bibnamefont {Bachmann}}, \bibinfo {author} {\bibfnamefont {J.~C.}\
  \bibnamefont {Toussaint}}, \bibinfo {author} {\bibfnamefont {O.}~\bibnamefont
  {Fruchart}},\ and\ \bibinfo {author} {\bibfnamefont {D.}~\bibnamefont
  {Gusakova}},\ }\emph {\bibinfo {title} {Mechanism of fast domain wall motion
  via current-assisted bloch-point domain wall stabilization}},\ \href
  {https://doi.org/10.1103/physrevb.103.054430} {\bibfield  {journal} {\bibinfo
   {journal} {\prb}\ }\textbf {\bibinfo {volume} {103}},\ \bibinfo {pages}
  {054430} (\bibinfo {year} {2021})}\BibitemShut {NoStop}%
\bibitem [{\citenamefont {Álvaro Gómez}\ \emph {et~al.}(2025)\citenamefont
  {Álvaro Gómez}, \citenamefont {Hurst}, \citenamefont {Hegde}, \citenamefont
  {Ruiz-Gómez}, \citenamefont {Pereiro}, \citenamefont {Aballe}, \citenamefont
  {Toussaint}, \citenamefont {Pérez}, \citenamefont {Masseboeuf},
  \citenamefont {Thirion}, \citenamefont {Fruchart},\ and\ \citenamefont
  {Gusakova}}]{bib-FRU2025}%
  \BibitemOpen
  \bibfield  {author} {\bibinfo {author} {\bibfnamefont {L.}~\bibnamefont
  {Álvaro Gómez}}, \bibinfo {author} {\bibfnamefont {J.}~\bibnamefont
  {Hurst}}, \bibinfo {author} {\bibfnamefont {S.}~\bibnamefont {Hegde}},
  \bibinfo {author} {\bibfnamefont {S.}~\bibnamefont {Ruiz-Gómez}}, \bibinfo
  {author} {\bibfnamefont {E.}~\bibnamefont {Pereiro}}, \bibinfo {author}
  {\bibfnamefont {L.}~\bibnamefont {Aballe}}, \bibinfo {author} {\bibfnamefont
  {J.~C.}\ \bibnamefont {Toussaint}}, \bibinfo {author} {\bibfnamefont
  {L.}~\bibnamefont {Pérez}}, \bibinfo {author} {\bibfnamefont
  {A.}~\bibnamefont {Masseboeuf}}, \bibinfo {author} {\bibfnamefont
  {C.}~\bibnamefont {Thirion}}, \bibinfo {author} {\bibfnamefont
  {O.}~\bibnamefont {Fruchart}},\ and\ \bibinfo {author} {\bibfnamefont
  {D.}~\bibnamefont {Gusakova}},\ }\emph {\bibinfo {title} {Topological
  analysis and experimental control of transformations of domain walls in
  magnetic cylindrical nanowires}},\ \href
  {https://doi.org/10.1103/physrevresearch.7.023092} {\bibfield  {journal}
  {\bibinfo  {journal} {\PhysRevRes}\ }\textbf {\bibinfo {volume} {7}},\
  \bibinfo {pages} {023092} (\bibinfo {year} {2025})}\BibitemShut {NoStop}%
\bibitem [{\citenamefont {{Van Waeyenberge}}\ \emph {et~al.}(2006)\citenamefont
  {{Van Waeyenberge}}, \citenamefont {Puzic}, \citenamefont {Stoll},
  \citenamefont {Chou}, \citenamefont {Tyliszczak}, \citenamefont {Hertel},
  \citenamefont {Fähnle}, \citenamefont {Brückl}, \citenamefont {Rott},
  \citenamefont {Reiss}, \citenamefont {Neudecker}, \citenamefont {Weiss},
  \citenamefont {Back},\ and\ \citenamefont {Schütz}}]{bib-VAN2006}%
  \BibitemOpen
  \bibfield  {author} {\bibinfo {author} {\bibfnamefont {B.}~\bibnamefont {{Van
  Waeyenberge}}}, \bibinfo {author} {\bibfnamefont {A.}~\bibnamefont {Puzic}},
  \bibinfo {author} {\bibfnamefont {H.}~\bibnamefont {Stoll}}, \bibinfo
  {author} {\bibfnamefont {K.~W.}\ \bibnamefont {Chou}}, \bibinfo {author}
  {\bibfnamefont {T.}~\bibnamefont {Tyliszczak}}, \bibinfo {author}
  {\bibfnamefont {R.}~\bibnamefont {Hertel}}, \bibinfo {author} {\bibfnamefont
  {M.}~\bibnamefont {Fähnle}}, \bibinfo {author} {\bibfnamefont
  {H.}~\bibnamefont {Brückl}}, \bibinfo {author} {\bibfnamefont
  {K.}~\bibnamefont {Rott}}, \bibinfo {author} {\bibfnamefont {G.}~\bibnamefont
  {Reiss}}, \bibinfo {author} {\bibfnamefont {I.}~\bibnamefont {Neudecker}},
  \bibinfo {author} {\bibfnamefont {D.}~\bibnamefont {Weiss}}, \bibinfo
  {author} {\bibfnamefont {C.~H.}\ \bibnamefont {Back}},\ and\ \bibinfo
  {author} {\bibfnamefont {G.}~\bibnamefont {Schütz}},\ }\emph {\bibinfo
  {title} {Magnetic vortex core reversal by excitation with short bursts of an
  alternating field}},\ \href {https://doi.org/10.1038/nature05240} {\bibfield
  {journal} {\bibinfo  {journal} {\nature}\ }\textbf {\bibinfo {volume}
  {444}},\ \bibinfo {pages} {461} (\bibinfo {year} {2006})}\BibitemShut
  {NoStop}%
\bibitem [{\citenamefont {Hertel}\ \emph {et~al.}(2007)\citenamefont {Hertel},
  \citenamefont {Gliga}, \citenamefont {Fähnle},\ and\ \citenamefont
  {Schneider}}]{bib-HER2007}%
  \BibitemOpen
  \bibfield  {author} {\bibinfo {author} {\bibfnamefont {R.}~\bibnamefont
  {Hertel}}, \bibinfo {author} {\bibfnamefont {S.}~\bibnamefont {Gliga}},
  \bibinfo {author} {\bibfnamefont {M.}~\bibnamefont {Fähnle}},\ and\ \bibinfo
  {author} {\bibfnamefont {C.~M.}\ \bibnamefont {Schneider}},\ }\emph {\bibinfo
  {title} {Nanomagnetic toggle switching of vortex cores on the picosecond time
  scale}},\ \href@noop {} {\bibfield  {journal} {\bibinfo  {journal} {\prl}\
  }\textbf {\bibinfo {volume} {98}},\ \bibinfo {pages} {117201} (\bibinfo
  {year} {2007})}\BibitemShut {NoStop}%
\bibitem [{\citenamefont {Yamada}\ \emph {et~al.}(2007)\citenamefont {Yamada},
  \citenamefont {Kasai}, \citenamefont {Nakatani}, \citenamefont {Kobayashi},
  \citenamefont {Kohno}, \citenamefont {Thiaville},\ and\ \citenamefont
  {Ono}}]{bib-YAM2007}%
  \BibitemOpen
  \bibfield  {author} {\bibinfo {author} {\bibfnamefont {K.}~\bibnamefont
  {Yamada}}, \bibinfo {author} {\bibfnamefont {S.}~\bibnamefont {Kasai}},
  \bibinfo {author} {\bibfnamefont {Y.}~\bibnamefont {Nakatani}}, \bibinfo
  {author} {\bibfnamefont {K.}~\bibnamefont {Kobayashi}}, \bibinfo {author}
  {\bibfnamefont {H.}~\bibnamefont {Kohno}}, \bibinfo {author} {\bibfnamefont
  {A.}~\bibnamefont {Thiaville}},\ and\ \bibinfo {author} {\bibfnamefont
  {T.}~\bibnamefont {Ono}},\ }\emph {\bibinfo {title} {Electrical switching of
  the vortex core in a magnetic disk}},\ \href
  {https://doi.org/10.1038/nmat1867} {\bibfield  {journal} {\bibinfo  {journal}
  {\NatMater}\ }\textbf {\bibinfo {volume} {6}},\ \bibinfo {pages} {270}
  (\bibinfo {year} {2007})}\BibitemShut {NoStop}%
\bibitem [{\citenamefont {Ot{\'{a}}lora}\ \emph {et~al.}(2017)\citenamefont
  {Ot{\'{a}}lora}, \citenamefont {Yan}, \citenamefont {Schultheiss},
  \citenamefont {Hertel},\ and\ \citenamefont {K{\'{a}}kay}}]{bib-OTA2017}%
  \BibitemOpen
  \bibfield  {author} {\bibinfo {author} {\bibfnamefont {J.~A.}\ \bibnamefont
  {Ot{\'{a}}lora}}, \bibinfo {author} {\bibfnamefont {M.}~\bibnamefont {Yan}},
  \bibinfo {author} {\bibfnamefont {H.}~\bibnamefont {Schultheiss}}, \bibinfo
  {author} {\bibfnamefont {R.}~\bibnamefont {Hertel}},\ and\ \bibinfo {author}
  {\bibfnamefont {A.}~\bibnamefont {K{\'{a}}kay}},\ }\emph {\bibinfo {title}
  {Asymmetric spin-wave dispersion in ferromagnetic nanotubes induced by
  surface curvature}},\ \href {https://doi.org/10.1103/physrevb.95.184415}
  {\bibfield  {journal} {\bibinfo  {journal} {\prb}\ }\textbf {\bibinfo
  {volume} {95}},\ \bibinfo {pages} {184415} (\bibinfo {year}
  {2017})}\BibitemShut {NoStop}%
\bibitem [{\citenamefont {Farinha}\ \emph {et~al.}(2025)\citenamefont
  {Farinha}, \citenamefont {Yang}, \citenamefont {Yoon}, \citenamefont {Pal},\
  and\ \citenamefont {Parkin}}]{bib-FAR2025}%
  \BibitemOpen
  \bibfield  {author} {\bibinfo {author} {\bibfnamefont {A.~M.~A.}\
  \bibnamefont {Farinha}}, \bibinfo {author} {\bibfnamefont {S.-H.}\
  \bibnamefont {Yang}}, \bibinfo {author} {\bibfnamefont {J.}~\bibnamefont
  {Yoon}}, \bibinfo {author} {\bibfnamefont {B.}~\bibnamefont {Pal}},\ and\
  \bibinfo {author} {\bibfnamefont {S.~S.~P.}\ \bibnamefont {Parkin}},\ }\emph
  {\bibinfo {title} {Interplay of geometrical and spin chiralities in
  3d twisted magnetic ribbons}},\ \href
  {https://doi.org/10.1038/s41586-024-08582-8} {\bibfield  {journal} {\bibinfo
  {journal} {Nature}\ }\textbf {\bibinfo {volume} {639}},\ \bibinfo {pages}
  {67} (\bibinfo {year} {2025})}\BibitemShut {NoStop}%
\bibitem [{\citenamefont {Schryer}\ and\ \citenamefont
  {Walker}(1974)}]{bib-SCH1974}%
  \BibitemOpen
  \bibfield  {author} {\bibinfo {author} {\bibfnamefont {N.~L.}\ \bibnamefont
  {Schryer}}\ and\ \bibinfo {author} {\bibfnamefont {L.~R.}\ \bibnamefont
  {Walker}},\ }\emph {\bibinfo {title} {The motion of 180° domain walls in
  uniform dc magnetic fields}},\ \href@noop {} {\bibfield  {journal} {\bibinfo
  {journal} {\jap}\ }\textbf {\bibinfo {volume} {45}},\ \bibinfo {pages} {5406}
  (\bibinfo {year} {1974})}\BibitemShut {NoStop}%
\bibitem [{\citenamefont {Lewis}\ \emph {et~al.}(2010)\citenamefont {Lewis},
  \citenamefont {Petit}, \citenamefont {O’Brien}, \citenamefont
  {Fernandez-Pacheco}, \citenamefont {Sampaio}, \citenamefont {Jausovec},
  \citenamefont {Read},\ and\ \citenamefont {Cowburn}}]{bib-LEW2010}%
  \BibitemOpen
  \bibfield  {author} {\bibinfo {author} {\bibfnamefont {E.~R.}\ \bibnamefont
  {Lewis}}, \bibinfo {author} {\bibfnamefont {D.}~\bibnamefont {Petit}},
  \bibinfo {author} {\bibfnamefont {L.}~\bibnamefont {O’Brien}}, \bibinfo
  {author} {\bibfnamefont {A.}~\bibnamefont {Fernandez-Pacheco}}, \bibinfo
  {author} {\bibfnamefont {J.}~\bibnamefont {Sampaio}}, \bibinfo {author}
  {\bibfnamefont {A.-V.}\ \bibnamefont {Jausovec}}, \bibinfo {author}
  {\bibfnamefont {H.~T. Z. D.~E.}\ \bibnamefont {Read}},\ and\ \bibinfo
  {author} {\bibfnamefont {R.~P.}\ \bibnamefont {Cowburn}},\ }\emph {\bibinfo
  {title} {Fast domain wall motion in magnetic comb structures}},\ \href@noop
  {} {\bibfield  {journal} {\bibinfo  {journal} {\NatMater}\ ,\ \bibinfo
  {pages} {980}} (\bibinfo {year} {2010})}\BibitemShut {NoStop}%
\bibitem [{bib({\natexlab{b}})}]{bib-CC-BY}%
  \BibitemOpen
  \href {https://creativecommons.org/licenses/by/4.0/} {} ({\natexlab{b}}),\
  \bibinfo {note}
  {\href{https://creativecommons.org/licenses/by/4.0/}{https://creativecommons.org/licenses/by/4.0/}}\BibitemShut
  {NoStop}%
\bibitem [{\citenamefont {Dom\'inguez-Bajo}\ \emph {et~al.}(2021)\citenamefont
  {Dom\'inguez-Bajo}, \citenamefont {Rosa}, \citenamefont {Gonz\'alez-Mayorga},
  \citenamefont {Rodilla}, \citenamefont {Arch\'e-Núñez}, \citenamefont
  {Benayas}, \citenamefont {Oc\'on}, \citenamefont {P\'erez}, \citenamefont
  {Camarero}, \citenamefont {Miranda}, \citenamefont {Gonz\'alez},
  \citenamefont {Aguilar}, \citenamefont {L\'opez-Dolado},\ and\ \citenamefont
  {Serrano}}]{bib-DOM2021}%
  \BibitemOpen
  \bibfield  {author} {\bibinfo {author} {\bibfnamefont {A.}~\bibnamefont
  {Dom\'inguez-Bajo}}, \bibinfo {author} {\bibfnamefont {J.~M.}\ \bibnamefont
  {Rosa}}, \bibinfo {author} {\bibfnamefont {A.}~\bibnamefont
  {Gonz\'alez-Mayorga}}, \bibinfo {author} {\bibfnamefont {B.~L.}\ \bibnamefont
  {Rodilla}}, \bibinfo {author} {\bibfnamefont {A.}~\bibnamefont
  {Arch\'e-Núñez}}, \bibinfo {author} {\bibfnamefont {E.}~\bibnamefont
  {Benayas}}, \bibinfo {author} {\bibfnamefont {P.}~\bibnamefont {Oc\'on}},
  \bibinfo {author} {\bibfnamefont {L.}~\bibnamefont {P\'erez}}, \bibinfo
  {author} {\bibfnamefont {J.}~\bibnamefont {Camarero}}, \bibinfo {author}
  {\bibfnamefont {R.}~\bibnamefont {Miranda}}, \bibinfo {author} {\bibfnamefont
  {M.~T.}\ \bibnamefont {Gonz\'alez}}, \bibinfo {author} {\bibfnamefont
  {J.}~\bibnamefont {Aguilar}}, \bibinfo {author} {\bibfnamefont
  {E.}~\bibnamefont {L\'opez-Dolado}},\ and\ \bibinfo {author} {\bibfnamefont
  {M.~C.}\ \bibnamefont {Serrano}},\ }\emph {\bibinfo {title} {Nanostructured
  gold electrodes promote neural maturation and network connectivity}},\ \href
  {https://doi.org/https://doi.org/10.1016/j.biomaterials.2021.121186}
  {\bibfield  {journal} {\bibinfo  {journal} {Biomaterials}\ }\textbf {\bibinfo
  {volume} {279}},\ \bibinfo {pages} {121186} (\bibinfo {year}
  {2021})}\BibitemShut {NoStop}%
\bibitem [{\citenamefont {Couderchon}\ and\ \citenamefont
  {Porteseil}(1996)}]{bib-COU1996}%
  \BibitemOpen
  \bibfield  {author} {\bibinfo {author} {\bibfnamefont {G.}~\bibnamefont
  {Couderchon}}\ and\ \bibinfo {author} {\bibfnamefont {J.~L.}\ \bibnamefont
  {Porteseil}},\ }\bibinfo {title} {The iron-nickel alloys}\ (\bibinfo
  {publisher} {Lavoisier},\ \bibinfo {year} {1996})\ Chap.\ \bibinfo {chapter}
  {Some properties of nickel-rich commercial Fe-Ni alloys}, p.~\bibinfo {pages}
  {27}\BibitemShut {NoStop}%
\bibitem [{\citenamefont {Lo}\ \emph {et~al.}(1987)\citenamefont {Lo},
  \citenamefont {Hwang}, \citenamefont {Huang}, \citenamefont {Campbell},\ and\
  \citenamefont {Allee}}]{bib-LO1987}%
  \BibitemOpen
  \bibfield  {author} {\bibinfo {author} {\bibfnamefont {J.}~\bibnamefont
  {Lo}}, \bibinfo {author} {\bibfnamefont {C.}~\bibnamefont {Hwang}}, \bibinfo
  {author} {\bibfnamefont {T.~C.}\ \bibnamefont {Huang}}, \bibinfo {author}
  {\bibfnamefont {R.}~\bibnamefont {Campbell}},\ and\ \bibinfo {author}
  {\bibfnamefont {D.}~\bibnamefont {Allee}},\ }\emph {\bibinfo {title}
  {Magnetic and structural properties of high rate dual ion-beam sputtered nife
  films (invited)}},\ \href {https://doi.org/10.1063/1.338713} {\bibfield
  {journal} {\bibinfo  {journal} {Journal of Applied Physics}\ }\textbf
  {\bibinfo {volume} {61}},\ \bibinfo {pages} {3520} (\bibinfo {year}
  {1987})}\BibitemShut {NoStop}%
\bibitem [{\citenamefont {Counil}\ \emph {et~al.}(2006)\citenamefont {Counil},
  \citenamefont {Devolder}, \citenamefont {Kim}, \citenamefont {Crozat},
  \citenamefont {Chappert}, \citenamefont {Zoll},\ and\ \citenamefont
  {Fournel}}]{bib-COU2006}%
  \BibitemOpen
  \bibfield  {author} {\bibinfo {author} {\bibfnamefont {G.}~\bibnamefont
  {Counil}}, \bibinfo {author} {\bibfnamefont {T.}~\bibnamefont {Devolder}},
  \bibinfo {author} {\bibfnamefont {J.-V.}\ \bibnamefont {Kim}}, \bibinfo
  {author} {\bibfnamefont {P.}~\bibnamefont {Crozat}}, \bibinfo {author}
  {\bibfnamefont {C.}~\bibnamefont {Chappert}}, \bibinfo {author}
  {\bibfnamefont {S.}~\bibnamefont {Zoll}},\ and\ \bibinfo {author}
  {\bibfnamefont {R.}~\bibnamefont {Fournel}},\ }\emph {\bibinfo {title}
  {Temperature dependences of the resistivity and the ferromagnetic resonance
  linewidth in permalloy thin films}},\ \href
  {https://doi.org/10.1109/tmag.2006.879718} {\bibfield  {journal} {\bibinfo
  {journal} {\IEEEtm}\ }\textbf {\bibinfo {volume} {42}},\ \bibinfo {pages}
  {3323} (\bibinfo {year} {2006})}\BibitemShut {NoStop}%
\bibitem [{\citenamefont {Neethirajan}\ \emph {et~al.}(2024)\citenamefont
  {Neethirajan}, \citenamefont {Daurer}, \citenamefont {Martínez},
  \citenamefont {Hrabec}, \citenamefont {Turnbull}, \citenamefont {Yamamoto},
  \citenamefont {Ferreira}, \citenamefont {Štefančič}, \citenamefont
  {Mayoh}, \citenamefont {Balakrishnan}, \citenamefont {Pei}, \citenamefont
  {Xue}, \citenamefont {Chang}, \citenamefont {Ringe}, \citenamefont
  {Harrison}, \citenamefont {Valencia}, \citenamefont {Kazemian}, \citenamefont
  {Kaulich},\ and\ \citenamefont {Donnelly}}]{bib-NEE2024}%
  \BibitemOpen
  \bibfield  {author} {\bibinfo {author} {\bibfnamefont {J.}~\bibnamefont
  {Neethirajan}}, \bibinfo {author} {\bibfnamefont {B.~J.}\ \bibnamefont
  {Daurer}}, \bibinfo {author} {\bibfnamefont {M.~D.~P.}\ \bibnamefont
  {Martínez}}, \bibinfo {author} {\bibfnamefont {A.}~\bibnamefont {Hrabec}},
  \bibinfo {author} {\bibfnamefont {L.}~\bibnamefont {Turnbull}}, \bibinfo
  {author} {\bibfnamefont {R.}~\bibnamefont {Yamamoto}}, \bibinfo {author}
  {\bibfnamefont {M.~R.}\ \bibnamefont {Ferreira}}, \bibinfo {author}
  {\bibfnamefont {A.}~\bibnamefont {Štefančič}}, \bibinfo {author}
  {\bibfnamefont {D.~A.}\ \bibnamefont {Mayoh}}, \bibinfo {author}
  {\bibfnamefont {G.}~\bibnamefont {Balakrishnan}}, \bibinfo {author}
  {\bibfnamefont {Z.}~\bibnamefont {Pei}}, \bibinfo {author} {\bibfnamefont
  {P.}~\bibnamefont {Xue}}, \bibinfo {author} {\bibfnamefont {L.}~\bibnamefont
  {Chang}}, \bibinfo {author} {\bibfnamefont {E.}~\bibnamefont {Ringe}},
  \bibinfo {author} {\bibfnamefont {R.}~\bibnamefont {Harrison}}, \bibinfo
  {author} {\bibfnamefont {S.}~\bibnamefont {Valencia}}, \bibinfo {author}
  {\bibfnamefont {M.}~\bibnamefont {Kazemian}}, \bibinfo {author}
  {\bibfnamefont {B.}~\bibnamefont {Kaulich}},\ and\ \bibinfo {author}
  {\bibfnamefont {C.}~\bibnamefont {Donnelly}},\ }\emph {\bibinfo {title} {Soft
  x-ray phase nanomicroscopy of micrometer-thick magnets}},\ \href
  {https://doi.org/10.1103/physrevx.14.031028} {\bibfield  {journal} {\bibinfo
  {journal} {\prx}\ }\textbf {\bibinfo {volume} {14}},\ \bibinfo {pages}
  {031028} (\bibinfo {year} {2024})}\BibitemShut {NoStop}%
\bibitem [{\citenamefont {Geuzaine}\ and\ \citenamefont
  {Remacle}(2009)}]{bib-GEU2009}%
  \BibitemOpen
  \bibfield  {author} {\bibinfo {author} {\bibfnamefont {C.}~\bibnamefont
  {Geuzaine}}\ and\ \bibinfo {author} {\bibfnamefont {J.}~\bibnamefont
  {Remacle}},\ }\emph {\bibinfo {title} {Gmsh: A 3‐d finite element mesh
  generator with built‐in pre‐ and post‐processing facilities}},\ \href
  {https://doi.org/10.1002/nme.2579} {\bibfield  {journal} {\bibinfo  {journal}
  {\ijnme}\ }\textbf {\bibinfo {volume} {79}},\ \bibinfo {pages} {1309}
  (\bibinfo {year} {2009})}\BibitemShut {NoStop}%
\end{thebibliography}

%

\end{document}